\documentclass[
	aps, prd, reprint,
	10pt, notitlepage, a4paper,
         floats, floatfix,
	amsmath, amssymb, amsfonts, eqsecnum,
	superscriptaddress,
	showpacs, showkeys,
	nofootinbib,
 	longbibliography,
]{revtex4-1}

\usepackage[usenames,dvipsnames]{xcolor}
\colorlet{LightRubineRed}{RubineRed!70}
\colorlet{Mycolor1}{green!10!orange}
\definecolor{Mycolor2}{HTML}{00F9DE}
\usepackage{amssymb}
\usepackage{amsmath}
\usepackage{cancel}
\usepackage{tikz}
\usepackage{caption}
\usepackage[compat=1.1.0]{tikz-feynman}
\usetikzlibrary[shapes.misc] 
\usetikzlibrary{arrows,snakes,backgrounds}
\usepackage{aas_macros}
\usepackage{todonotes}
\usepackage{xspace} 
\usepackage{bm,graphicx} 
\usepackage[utf8]{inputenc} 
\newcommand{\be}{\begin{equation}}
\newcommand{\ee}{\end{equation}}
\newcommand{\bea}{\begin{eqnarray}}
\newcommand{\eea}{\end{eqnarray}}

\newcommand{\bdm}{\begin{displaymath}}
\newcommand{\edm}{\end{displaymath}}

\xdefinecolor{mylinkcolor}{rgb}{0,0,0.5}
\usepackage[
	bookmarksnumbered, bookmarksopen, bookmarksopenlevel=2,
	breaklinks=true,
	colorlinks=true, filecolor=mylinkcolor, citecolor=mylinkcolor,
	linkcolor=mylinkcolor, urlcolor=mylinkcolor, menucolor=mylinkcolor,
]{hyperref}

\begin{document}

\title{Binary Black Holes in Einstein-Maxwell-Dilaton Theory: Second Post-Newtonian Dynamics from Effective Field Theory}

\author{Pawan Kumar Gupta}
\email{pkgupta@camk.edu.pl}
\affiliation{Nicolaus Copernicus Astronomical Center, Polish Academy of Sciences, 00-716, Warsaw, Poland}
\begin{abstract}
The detection of gravitational waves from compact binary coalescences provides a powerful opportunity to test general relativity in the strong-field regime and to search for signatures of alternative theories of gravity. In this work, we consider Einstein-Maxwell-Dilaton (EMd) theory, in which black holes can carry both electric and scalar (dilatonic) charges. We employ the effective field theory approach, together with a temporal Kaluza-Klein decomposition of the metric in terms of non-relativistic gravitational fields, to derive the conservative two-body Lagrangian for charged black-hole binaries in EMd theory through second post-Newtonian (PN) order. Our calculation extends the previously known conservative dynamics at 1PN order and includes the gravitational, electromagnetic, and dilaton interactions at 2PN order. We verify the result in the appropriate Einstein-Maxwell, scalar-tensor, and general relativistic limits, and perform an independent test-body-limit check of the static 2PN sector. These results provide the conservative dynamics needed for developing higher-accuracy waveform models and testing EMd theory with gravitational-wave observations.

\end{abstract}
\maketitle

\section{Introduction}
The observation of gravitational waves (GWs) from compact binary coalescences by Advanced LIGO~\cite{LIGOScientific:2014pky}, Advanced Virgo~\cite{VIRGO:2014yos} and
KAGRA~\cite{KAGRA:2018plz} has opened unprecedented opportunities for probing general relativity (GR) in the strong-field, highly dynamical regime~\cite{Will:2014kxa, Berti:2015itd, Yunes:2016jcc, LIGOScientific:2016lio, LIGOScientific:2018dkp, LIGOScientific:2020tif}. Realizing this potential requires accurate waveform models that incorporate possible deviations from GR. Such deviations can be parametrized in a theory-agnostic manner and constrained observationally~\cite{Arun:2006hn,Yunes:2009ke,Mishra:2010tp,Li:2011cg,Shoom:2021mdj}, as implemented by the LIGO-Virgo-KAGRA collaborations~\cite{LIGOScientific:2019fpa,LIGOScientific:2021sio,LIGOScientific:2021nrg,LIGOScientific:2026fcf,LIGOScientific:2026gwtc5gr}. Alternatively, waveform models derived within specific modified gravity theories allow for direct constraints on the underlying parameters of these theories~\cite{Gupta:2021rod,Yagi:2011xp,Yagi:2011ze,Yagi:2012vf,Yunes:2011we,Yagi:2012ya,Silva:2017uqg,Roy:2025xih}. Such frameworks provide a means to probe extensions of GR, including additional long-range scalar degrees of freedom coupled to compact objects~\cite{Berti:2018cxi,Sotiriou:2015pka}. In many modified gravity theories, these extra degrees of freedom manifest as scalar or vector hair, endowing compact objects with additional charges and leading to observable modifications in GW signals~\cite{Sotiriou:2015pka,Silva:2017uqg,vanGemeren:2023rhh}.

A well-motivated such theory is EMd theory, in which gravity is coupled to both an electromagnetic field and a scalar dilaton. Such theories arise naturally in the low-energy limit of string theory~\cite{Garfinkle:1990qj,Gibbons:1987ps} and in Kaluza-Klein compactifications, where the dilaton governs the effective gauge coupling. EMd theory admits analytic black hole solutions characterized by a dilaton coupling $a$, with $a=0$ recovering Reissner-Nordström, $a=1$ corresponding to heterotic string theory, and $a=\sqrt{3}$ to Kaluza-Klein black holes~\cite{Horne:1992zy,Frolov:1987rj}. These black holes can carry both electric and scalar (dilatonic) charges, and in the absence of these charges, the black-hole solutions reduce to their GR counterparts. 

In GR (and some scalar extensions), isolated black holes cannot support additional charges, as encapsulated by no-hair theorems~\cite{Hawking:1972qk,Sotiriou:2011dz}. Moreover, charged black holes are expected to neutralize efficiently via pair production or interactions with surrounding plasma, severely limiting their charge within the Standard Model due to the small electron mass-to-charge ratio~\cite{Gibbons:1975kk,Eardley:1975kp}. Beyond the Standard Model, however, scenarios such as minicharged or hidden-sector dark matter can significantly increase the effective mass-to-charge ratio, allowing black holes to acquire and retain substantial (visible or hidden) charge while remaining consistent with current experimental and cosmological constraints~\cite{Cardoso:2016olt,Feng:2009mn,Kadota:2016tqq}.

The dynamics of black-hole binaries in EMd theory has been investigated using both numerical and analytical approaches. Numerical-relativity simulations of isolated and binary black holes with small electric charges have shown that the resulting GW signals can be difficult to distinguish from their GR counterparts~\cite{Hirschmann:2017psw}. On the analytical side, the inspiral dynamics and the corresponding modifications to the GW signal have been studied within the post-Newtonian (PN) framework up to 1PN order~\cite{Julie:2017rpw,Khalil:2018aaj}. The leading modification to the GW phase arises from dipolar radiation, which enters at $-1$PN order relative to the leading quadrupolar radiation,
while additional corrections enter at higher PN orders through modifications
to both the conservative and radiative dynamics. 

The conservative sector of the binary dynamics is a key ingredient in GW modeling, entering both the construction of PN inspiral waveforms~\cite{Blanchet:2013haa} and the development of inspiral-merger-ringdown models within frameworks such as the effective-one-body approach~\cite{Buonanno:1998gg,Buonanno:2000ef}. Extending the conservative dynamics of EMd binaries to higher PN orders is therefore an important step toward developing more accurate waveform models and testing the theory with GW observations.

In this paper, we focus on the conservative two-body dynamics of charged black-hole binaries in EMd theory through second post-Newtonian (2PN) order. We perform the calculation using the effective field theory (EFT) approach to the relativistic two-body problem, originally developed by Goldberger and Rothstein~\cite{Goldberger:2004jt}, extending the non-relativistic general relativity (NRGR) framework to include the electromagnetic and scalar degrees of freedom of EMd theory. A central ingredient of our analysis is the use of the non-relativistic gravitational (NRG) field decomposition introduced by Kol and Smolkin~\cite{Kol:2007bc}. This parametrization exploits the hierarchy of scales inherent in the PN expansion and greatly reduces the number and complexity of Feynman diagrams. In the presence of electromagnetic and dilaton fields, the NRG formalism reveals a clear coupling hierarchy between gravitational, gauge, and scalar interactions, making it particularly well suited for EMd systems. 

At leading order, we recover the Newtonian gravitational potential supplemented by Coulomb and scalar-mediated interactions. At first post-Newtonian order, we reproduce known EMd results using a minimal set of diagrams. At second post-Newtonian order, we evaluate all relevant one-loop and two-loop contributions, carefully treating divergent integrals using dimensional regularization. Our results provide a complete and systematic description of the conservative dynamics of EMd binaries at 2PN accuracy.

This paper is organized as follows. In Sec.~\ref{sec:singleBH}, we introduce EMd theory and the effective point-particle description of compact objects, including the scalar-dependent mass and the associated scalar response parameters. In Sec.~\ref{sec:EFT FRAMEWORK}, we formulate the EFT description of the binary dynamics, introduce the NRG decomposition and the potential-mode expansion, and derive the worldline and bulk interaction vertices and propagators required for the PN calculation. We also discuss the power counting used to organize the diagrammatic expansion. In Sec.~\ref{section:0PN}, we compute the leading-order scalar potential using EFT methods. The 1PN and 2PN contributions to the conservative dynamics are derived in Secs.~\ref{section:0PN} and~\ref{sec:2PN}, respectively. In Sec.~\ref{sec:results}, we combine these contributions and present the complete conservative two-body Lagrangian through 2PN order, and perform consistency checks of the result, including an independent static test-body-limit check against the exact EMd black-hole solution. We conclude in Sec.~\ref{sec:conclusions} with a summary and outlook. Additional details of the test-body calculation and useful formulas are provided in Appendices~\ref{app:test-body-derivation} and~\ref{app:useful-formulas}.

Throughout this paper, we use the shorthand $\int_{\bm k}\equiv\int d^3{\bm k}/(2\pi)^3$. We set the speed of light $c=1$, except when explicitly displaying powers of $1/c^2$ to indicate the post-Newtonian (PN) order. Greek indices $\mu,\nu,\ldots$ denote four-dimensional spacetime indices, while lowercase Latin indices $i,j,k,\ldots$ denote spatial indices. We adopt the metric signature $\eta_{\mu\nu}=\mathrm{diag}(-1,+1,+1,+1)$ and the convention
\begin{equation}
R^{\mu}{}_{\nu\alpha\beta} = \Gamma^{\mu}{}_{\nu\beta,\alpha} -\Gamma^{\mu}{}_{\nu\alpha,\beta} +\Gamma^{\rho}{}_{\nu\beta}\Gamma^{\mu}{}_{\rho\alpha} -\Gamma^{\rho}{}_{\nu\alpha}\Gamma^{\mu}{}_{\rho\beta},
\end{equation}
for the Riemann tensor, where $\Gamma^{\mu}{}_{\nu\alpha}$ are the Christoffel symbols and a comma denotes a partial derivative.

The label $A=1,2$ denotes the two bodies in the binary. Their positions, velocities, accelerations, masses, and electric charges are denoted by $\bm{x}_A$, $\bm{v}_A$, $\bm{a}_A$, $m_A$, and $q_A$, respectively. The quantities $\alpha_A$, $\beta_A$, and $\beta'_A$ characterize their scalar couplings and nonlinear scalar response. We define the relative separation vector as $\bm{r}\equiv\bm{x}_1-\bm{x}_2$, with $r\equiv|\bm{r}|$, and the corresponding unit vector as $\bm{n}\equiv\bm{r}/r$.

\section{Einstein-Maxwell-Dilaton theory}
\label{sec:singleBH}
The Einstein frame provides a convenient formulation of EMd theory, which we adopt here. The equivalence between the Einstein and Jordan frames is discussed in Ref.~\cite{Flanagan:2004bz}. The action of EMd theory in the Einstein frame takes the form,
\begin{align}
\label{EMdaction}
\mathcal{S} =  \mathcal{S}_{g} + \mathcal{S}_{em} + \mathcal{S}_{\varphi} + \mathcal{S}_{m},
\end{align}
 
where $\mathcal{S}_g$ is the Einstein-Hilbert action supplemented by a harmonic gauge-fixing term:
\begin{equation}
\label{eq:sg}
    \mathcal{S}_{g} =  \frac{1}{16 \pi G} \int d^4 x \sqrt{-g} R -  \frac{1}{32 \pi G} \int d^4x\sqrt{-g}g_{\mu \nu} \Gamma^{\mu} \Gamma^{\nu},
\end{equation}
where $\Gamma^{\mu} \equiv \Gamma^{\mu}_{\rho \sigma} g^{\rho \sigma}$. The Maxwell-dilaton sector in curved spacetime, including the Feynman gauge-fixing term for the vector field, is
\begin{flalign}
    \label{eq:sqaction}
	\mathcal{S}_{em} = \frac{-1}{16\pi} \int d^4 x \sqrt{-g} \; e^{-2a\varphi}\left(  F^2  + 2(\nabla_{\mu} A^\mu)^2 \right),
\end{flalign}
where $F_{\mu\nu} = \nabla_{\mu} A_{\nu} - \nabla_{\nu}A_{\mu}$. The scalar-field action is
\begin{flalign}
    \label{eq:sqaction}
	\mathcal{S}_{\varphi} = \frac{-1}{16\pi G} \int d^4 x \sqrt{-g} \left( 2g^{\mu\nu}\partial_\mu\varphi\partial_\nu\varphi \right),
\end{flalign}

To model compact objects, we follow the approach of Eardley~\cite{eardley1975observable}, treating each body as a point particle with scalar-field-dependent mass. Neglecting spin, dipole moments, and higher multipoles, the matter action for charged monopolar particles can be written as~\cite{Damour1992Tensor}
\begin{equation}\label{Smatter}
\mathcal{S}_m = -\sum_A \int dt \left[ \mathfrak{m}_A(\varphi) \sqrt{-g_{\mu\nu} v_A^\mu v_A^\nu} - q_A A_\mu v_A^\mu \right],
\end{equation}
where $\mathfrak{m}_A(\varphi)$ is the field-dependent mass, $q_A$ is the electric/dark charge, and $v_A^\mu$ is the four-velocity of body $A$.
Since a closed-form expression for $\mathfrak{m}(\varphi)$ is typically unavailable, it is convenient to expand it about a background value $\varphi_0$ of the scalar field. Introducing $\delta\varphi \equiv \varphi - \varphi_0$, the expansion can be parameterized in terms of
\begin{equation}
\alpha(\varphi) \equiv \frac{d \ln \mathfrak{m}}{d\varphi}, 
\qquad 
\beta(\varphi) \equiv \frac{d\alpha}{d\varphi},
\qquad 
\beta'(\varphi) \equiv \frac{d\beta}{d\varphi},
\end{equation}
Here, $\alpha$ characterizes the scalar charge of the compact object, while $\beta$ and $\beta'$ characterize its nonlinear response to the scalar field. The mass then takes the form
\begin{align}
\label{Mphi}
\mathfrak{m}(\varphi)
= m \Bigg[ 1 &+ \alpha\,\delta\varphi + \frac{1}{2}(\alpha^2+\beta)\,\delta\varphi^2 \nonumber\\ &+ \frac{1}{3!} \left[ \alpha(\alpha^2+3\beta)+\beta' \right] \delta\varphi^3 + \mathcal{O}(\delta\varphi^4) \Bigg].
\end{align}

with $m \equiv \mathfrak{m}(\varphi_0)$. For notational simplicity, we denote $\alpha \equiv \alpha(\varphi_0)$, $\beta \equiv \beta(\varphi_0)$ and $\beta' \equiv \beta'(\varphi_0)$ unless otherwise specified. We set the background value of the scalar field to zero, \(\varphi_0 = 0\) without loss of generality~\cite{Khalil:2018aaj}.

In general, the scalar response depends on the internal structure of the compact object and is characterized by the scalar charge $\alpha$, which governs the linear coupling to the scalar field, and by $\beta$ and $\beta’$, which characterize the nonlinear scalar response. For black holes, however, these quantities are determined solely by the charge-to-mass ratio, whereas for ordinary matter they can also depend on the composition of the body. In EMd theory, the black-hole solution determines the scalar-field dependence of $\mathfrak m(\varphi)$ and hence fixes the corresponding functions $\alpha(\varphi)$, $\beta(\varphi)$, and $\beta’(\varphi)$. For the special case $a=1$, corresponding to heterotic string theory, the differential equation for $\mathfrak m(\varphi)$ admits an exact analytic solution, yielding~\cite{Julie:2017rpw}
\begin{equation}
\begin{aligned}
\alpha &= \frac{q^2}{2Gm^2},
\qquad
\beta = \frac{q^2}{Gm^2}- \frac{q^4}{2G^2m^4},\\
\beta' &= \frac{2q^2}{Gm^2}- \frac{3q^4}{G^2m^4}+ \frac{q^6}{G^3m^6}.
\end{aligned}
\end{equation}
where $m\equiv\mathfrak m(\varphi_0)$ denotes the mass evaluated at the background scalar field $\varphi_0$, and $q$ is the black-hole charge.

\section{EFT FRAMEWORK}
\label{sec:EFT FRAMEWORK}
We work within the EFT framework of NRGR, which exploits the hierarchy of scales in the inspiral regime,
\(r_c \ll r \ll \lambda\), where \(r_c\) denotes the size of the compact objects, \(r\) the orbital separation, and \(\lambda\) the characteristic wavelength of the emitted radiation. This separation of scales enables a systematic expansion in the post-Newtonian parameter \(v/c\) and the decomposition of the dynamics into near- and far-zone contributions.

In this framework, the binary constituents are described as point-particle worldlines endowed with effective couplings, while the gravitational, electromagnetic, and scalar fields are treated as dynamical degrees of freedom. Finite-size effects can be incorporated through higher-dimensional worldline operators; in this work, we restrict to minimally coupled point charges and focus on the conservative dynamics up to 2PN order.

The inspiral dynamics of binary black holes in EMd theory is governed by the action~\eqref{EMdaction}. To implement the post-Newtonian expansion, we parametrize the metric using the Kaluza-Klein (KK) ansatz,
\begin{equation} \label{eq:kka}
ds^2 = g_{\mu\nu}dx^{\mu}dx^{\nu} \equiv -e^{2 \phi}(dt - \mathcal{A}_i\, dx^i)^2 + e^{-2 \phi} \gamma_{ij}dx^i dx^j~,
\end{equation}
where \(\{\phi, \mathcal{A}_i, \gamma_{ij}\equiv\delta_{ij}+\sigma_{ij}\}\) denote the non-relativistic gravitational (NRG) fields. The metric can then be written in matrix form as
\begin{equation}
    g_{\mu \nu} =
    \begin{pmatrix}
    -e^{2 \phi} & e^{2\phi} \mathcal{A}_j \\
    e^{2\phi} \mathcal{A}_{i} & e^{-2 \phi} \gamma_{ij} - e^{2 \phi} \mathcal{A}_{i} \mathcal{A}_{j}
    \end{pmatrix}.
    \label{eq:metricKK}
\end{equation}
The NRG fields admit a simple physical interpretation: \(\phi\) corresponds to the Newtonian potential, \(\mathcal{A}_{i}\) to the gravito-magnetic potential, and \(\sigma_{ij}\) to spatial metric perturbations~\cite{Kol:2010si, Kol:2007bc}. Their leading PN scalings are \(\mathcal{O}(1/c^2)\), \(\mathcal{O}(1/c^3)\), and \(\mathcal{O}(1/c^4)\), respectively.

Substituting this decomposition into the action~(2.1), we expand perturbatively in the weak-field, slow-motion regime and extract the interaction vertices required for the PN calculation.

To implement the EFT description, we decompose the fields into potential (near-zone) and radiation (far-zone) modes. The potential modes, characterized by $|\mathbf{k}|\sim1/r$ and $k^0\sim v/r$,
mediate the conservative interactions and are instantaneous at leading order in the PN expansion. Radiation modes, with $k^0\sim|\mathbf{k}|\sim v/r$, describe propagating waves and give rise to dissipative effects. An analogous decomposition is applied to the electromagnetic and dilaton fields.

At the 2PN order considered here, the conservative dynamics is determined by the near-zone potential modes. We therefore restrict the present calculation to potential modes. Their propagators, together with the interaction vertices obtained from the expanded action, form the building blocks of the diagrammatic calculations presented below.

\subsection{The matter action}

Substituting the mass expansion~\eqref{Mphi} into the matter action~\eqref{Smatter}, we decompose it into three distinct contributions,
\begin{equation}
\mathcal{S}_m = \mathcal{S}_{pp} + \mathcal{S}_{ppq} + \mathcal{S}_{pp\varphi},
\end{equation}
corresponding respectively to the purely gravitational coupling, the coupling to the electromagnetic field, and the interaction with the scalar (dilaton) field.
The first term,
\begin{equation}
\mathcal{S}_{pp} = -\sum_A \int dt \; m_A \sqrt{-g_{\mu\nu} v_A^\mu v_A^\nu},
\end{equation}
describes the minimal coupling of the compact objects to the spacetime metric and encodes the standard geodesic motion in the absence of additional fields.
The second term,
\begin{equation}
\mathcal{S}_{ppq} = \sum_A \int dt \; q_A A_\mu v_A^\mu,
\end{equation}
represents the interaction of the worldline charges with the electromagnetic field, leading to both Coulomb-type interactions and velocity-dependent magnetic effects.
The third contribution,
\begin{align}\label{Sppvarphi}
\mathcal{S}_{pp\varphi} &= -\sum_A \int dt \; m_A  \left[ \alpha_A\,\varphi_A + \frac{1}{2}(\alpha_A^2 + \beta_A)\,\varphi_A^2  \right. \nonumber \\
& \left. +\frac{1}{3!} \left[ \alpha_A(\alpha_A^2+3\beta_A)+\beta'_A \right] \right] \varphi^3 \times \sqrt{-g_{\mu\nu} v_A^\mu v_A^\nu}.
\end{align}
Here we have used $\varphi_0=0$, as specified in Sec.~II, such that
$\delta\varphi=\varphi$.

\subsubsection{Point Particle Action}
\label{sec:Point Particle Action}
In the EFT framework, compact objects are modeled as point particles propagating along worldlines. Their minimal coupling to gravity is described by the point-particle action
\begin{equation}
    \label{eq:Spp}
    \mathcal{S}_{pp} = -\sum_A \int dt \; m_A \sqrt{-g_{\mu\nu} v_A^{\mu} v_A^{\nu}},
\end{equation}
where \(m_A\) denotes the mass of each body. We parametrize the worldlines using the coordinate time \(t\).

Since both bodies contribute identically up to relabeling, it is sufficient to consider a single worldline and restore the sum over particles at the end. To implement the post-Newtonian expansion, we substitute the Kaluza-Klein metric~\eqref{eq:kka} into the action and expand in powers of the fields and velocities.

\begin{flalign}
\begin{split}
    \mathcal{S}_{pp} &= -m \int dt \left( 1 - \frac{1}{2}{\bm v^2} + \phi  - \mathcal{A}_iv^i - \frac{1}{8}v^4 + \frac{3}{2}\phi v^2 + \frac{1}{2}\phi^2   \right. \\ & \left. - \frac{1}{2} v^i v^j \sigma_{ij}   -\frac{1}{16}v^6 \right),
\end{split}
\label{eq:sppkk}
\end{flalign}

where we have used the leading PN scaling of the NRG fields to consistently retain terms up to 2PN order. From this expansion, we extract the worldline-field interaction vertices corresponding to the coupling of the NRG fields to the particle mass:
\begin{align}
    \label{gravitonscalarvertex}
        \begin{gathered}
   \begin{tikzpicture}
\draw[black, ultra thick] (0,-1.5) -- (0,1.5);
\draw[black,  thick] (0,0) -- (1.5,0);
\filldraw[black] (0,0) circle (3pt) node[anchor=west]{};
\end{tikzpicture}
        \end{gathered}  &= - m \int d t \phi \left\{ 1 + \frac{3}{2}v^2  \right\},
\end{align}
\begin{align}
    \label{gravitonvectorvertex}    
        \begin{gathered}
        \begin{tikzpicture}
        \draw[black, ultra thick] (0,-1.5) -- (0,1.5);
        \draw[snake=coil,segment aspect=0, thick] (0,0) -- (1.5,0);
        \filldraw[black] (0,0) circle (3pt) node[anchor=west]{};
        \end{tikzpicture}   
        \end{gathered} &= ~m \int dt~ \mathcal{A}_iv^i,
\end{align}
\begin{align}
    \label{gravitontensorvertex}     
        \begin{gathered}
        \begin{tikzpicture}
        \draw[black, ultra thick] (0,-1.5) -- (0,1.5);
        \draw[snake=coil,segment aspect=0, thick] (0,0) -- (1.5,0);
        \draw[snake=coil,segment aspect=0, thick] (0,0.1) -- (1.5,0.1);
        \filldraw[black] (0,0) circle (3pt) node[anchor=west]{};
        \end{tikzpicture}   
        \end{gathered}&= ~\frac{m}{2} \int dt~\sigma_{ij}v^iv^j~,
\end{align}
\begin{align}
    \begin{gathered}
   \begin{tikzpicture}
\draw[black, ultra thick] (0,-1.5) -- (0,1.5);
\draw[black,  thick] (0,0) -- (1,1);
\draw[black,  thick] (0,0) -- (1,-1);
\filldraw[black] (0,0) circle (3pt) node[anchor=west]{};
\end{tikzpicture}
    \end{gathered}
    ={}& - \frac{1}{2} m \int dt  \phi^2 .
\end{align}
In the diagrams, the solid vertical lines denote the particle worldlines, the filled circles represent the compact objects on the worldlines, and the attached lines denote their interactions with the corresponding NRG fields $\phi$, $\mathcal{A}_i$, and $\sigma_{ij}$.

\subsubsection{Charged Particle Action}
\label{sec:Charged Particle Action}
The contribution to the worldline action arising from the charge is
\begin{align}
	\mathcal{S}_{ppq} = \sum_A \int dt \; q_A A_\mu v_A^\mu = \sum_A \int d t \left(q_A A_{0} + q_A v^{i} A_{i} \right).
	\label{eq:chargedpp}
\end{align}
From this expression, we extract the worldline-field interaction vertices associated with the electromagnetic field. The coupling to the scalar component \(A_0\) yields
\begin{align}
    \label{emscalar}
    \begin{gathered}
   \begin{tikzpicture}
\draw[black, ultra thick] (0,-1.5) -- (0,1.5);
\draw[dashed, thick] (0,0) -- (1.5,0);
\filldraw[black] (0,0) circle (3pt) node[anchor=west]{};
\end{tikzpicture}
    \end{gathered}
    ={}& q \int dt  A_0.
\end{align}
\noindent Similarly, the coupling to the spatial components \(A_i\) gives:
\begin{align}
    \label{emvector}
    \begin{gathered}
   \begin{tikzpicture}
\draw[black, ultra thick] (0,-1.5) -- (0,1.5);
\draw[dashed, thick] (0,0.05) -- (1.5,0.05);
\draw[dashed, thick] (0,-0.05) -- (1.5,-0.05);
\filldraw[black] (0,0) circle (3pt) node[anchor=west]{};
\end{tikzpicture}
    \end{gathered}
    ={}& q \int dt  A_i v^i.
\end{align}

\subsubsection{Dilaton particle action}
\label{sec:DilatonParticle}

The contribution to the worldline action arising from the scalar (dilaton) field is
\begin{align}\label{Sppvarphi}
\mathcal{S}_{pp\varphi} &= -\sum_A \int dt \; m_A  \left[ \alpha_A\,\varphi_A + \frac{1}{2}(\alpha_A^2 + \beta_A)\,\varphi_A^2  \right. \nonumber \\
& \left. +\frac{1}{3!} \left[ \alpha_A(\alpha_A^2+3\beta_A)+\beta'_A \right] \right] \varphi^3 \times \sqrt{-g_{\mu\nu} v_A^\mu v_A^\nu}.
\end{align}
From this expression, we extract the worldline-field interaction vertices associated with the scalar field:
\begin{align}
    \label{emscalar}
    \begin{gathered}
   \begin{tikzpicture}
\draw[black, ultra thick] (0,-1.5) -- (0,1.5);
\draw[snake=zigzag,segment aspect=0,  thick] (0,0) -- (1.5,0);
\filldraw[black] (0,0) circle (3pt) node[anchor=west]{};
\end{tikzpicture}
    \end{gathered}
    ={}&  -\int dt  \;m\;  \alpha \; \varphi \left(1  -  \frac{1}{2}v^2 \right),
    \end{align}
\begin{align}
    \label{emscalar1}
    \begin{gathered}
   \begin{tikzpicture}
\draw[black, ultra thick] (0,-1.5) -- (0,1.5);
\draw[black,  thick] (0,0) -- (1,1);
\draw[snake=zigzag,segment aspect=0,  thick] (0,0) -- (1,-1);
\filldraw[black] (0,0) circle (3pt) node[anchor=west]{};
\end{tikzpicture}
    \end{gathered}
    ={}&  -\int dt \; m \; \alpha \; \varphi \; \phi,
    \end{align}
\begin{align}
    \label{emscalar2}
    \begin{gathered}
   \begin{tikzpicture}
\draw[black, ultra thick] (0,-1.5) -- (0,1.5);
\draw[snake=coil,segment aspect=0, thick] (0,0) -- (1,1);
\draw[snake=zigzag,segment aspect=0,  thick] (0,0) -- (1,-1);
\filldraw[black] (0,0) circle (3pt) node[anchor=west]{};
\end{tikzpicture}
    \end{gathered}
    ={}&  \int dt \; m \; \alpha \; \varphi \; \mathcal{A}_iv^i,
    \end{align}    
\begin{align}
    \label{emscalar3}
    \begin{gathered}
   \begin{tikzpicture}
\draw[black, ultra thick] (0,-1.5) -- (0,1.5);
\draw[snake=zigzag,segment aspect=0,  thick] (0,0) -- (1,1);
\draw[snake=zigzag,segment aspect=0,  thick] (0,0) -- (1,-1);
\filldraw[black] (0,0) circle (3pt) node[anchor=west]{};
\end{tikzpicture}
    \end{gathered}
    ={}&  -\int dt \; m \; \frac{1}{2}(\alpha^2+\beta) \; \varphi^2,
    \end{align}
\begin{align}
    \label{emscalar4}
    \begin{gathered}
   \begin{tikzpicture}
\draw[black, ultra thick] (0,-1.5) -- (0,1.5);
\draw[black,  thick] (0,0) -- (1,1);
\draw[black,  thick] (0,0) -- (1,-1);
\draw[snake=zigzag,segment aspect=0,  thick] (0,0) -- (1,0);
\filldraw[black] (0,0) circle (3pt) node[anchor=west]{};
\end{tikzpicture}
    \end{gathered}
    =-{}&  \int dt \; m \; \frac{1}{2} \; \alpha \; \varphi \; \phi^2.
    \end{align}
\begin{align}
    \label{emscalar5}
    \begin{gathered}
   \begin{tikzpicture}
\draw[black, ultra thick] (0,-1.5) -- (0,1.5);
\draw[black,  thick] (0,0) -- (1,1);
\draw[snake=zigzag,segment aspect=0,  thick] (0,0) -- (1,-1);
\draw[snake=zigzag,segment aspect=0,  thick] (0,0) -- (1,0);
\filldraw[black] (0,0) circle (3pt) node[anchor=west]{};
\end{tikzpicture}
    \end{gathered}
    =-{}&  \int dt \; m \; \frac{1}{2}(\alpha^2+\beta) \; \varphi^2 \phi.
    \end{align}        
\begin{align}
    \label{emscalar6}
    \begin{gathered}
   \begin{tikzpicture}
\draw[black, ultra thick] (0,-1.5) -- (0,1.5);
\draw[snake=zigzag,segment aspect=0,  thick] (0,0) -- (1,1);
\draw[snake=zigzag,segment aspect=0,  thick] (0,0) -- (1,-1);
\draw[snake=zigzag,segment aspect=0,  thick] (0,0) -- (1,0);
\filldraw[black] (0,0) circle (3pt) node[anchor=west]{};
\end{tikzpicture}
    \end{gathered}
    =-{}&  \int dt \; m \; \frac{1}{3!} \left[ \alpha_A(\alpha_A^2+3\beta_A)+\beta'_A \right] \varphi^3
\end{align}        
    
\subsection{Einstein-Hilbert Action}
\label{sec:Einstein-Hilbert}
We consider the gravitational action to consist of the Einstein-Hilbert action along with a harmonic gauge fixing term:
\begin{equation}
\label{eq:sg}
    \mathcal{S}_{g} = \frac{1}{16 \pi G} \int d^4 x \sqrt{-g} R -  \frac{1}{32 \pi G} \int d^4x\sqrt{-g}g_{\mu \nu} \Gamma^{\mu} \Gamma^{\nu},
\end{equation}
Substituting the KK metric~\eqref{eq:kka} into the gravitational action, one obtains~\cite{Kol:2007bc}:
\begin{flalign}
\mathcal{S}_g = \frac{1}{16 \pi G} \int d t d^3  x \sqrt{\gamma} \left( R[\gamma] - 2 \partial_{i} \phi \partial^{i} \phi + \frac{1}{4} e^{4 \phi} \mathcal{F}^2 \right),
\label{eq:KKgravity}
\end{flalign}
where $\partial_i\phi\,\partial^i\phi =\gamma^{ij}\partial_i\phi\,\partial_j\phi$ and $\mathcal{F}_{ij}\equiv\partial_i\mathcal{A}_j- \partial_j\mathcal{A}_i$ is the field strength associated with the
gravito-magnetic field $\mathcal{A}_i$.

From the quadratic part of the action~\eqref{eq:KKgravity}, we obtain the propagators for the NRG fields~\cite{Levi:2008nh}:
\begin{align}
    \label{scalargraviton}
    \left< \phi ({\bm x_1}, t_1) \phi ({\bm x_2},t_2)\right> &= 4 \pi G \; \delta(t_1 - t_2)  \int_{\bm k}  \frac{e^{i {\bm k}\cdot {\bm r}}}{{\bm k^2}}, \\
    \left< \mathcal{A}_i ({\bm x_1},t_1) \mathcal{A}_j({\bm x_2},t_2)\right> &= - 16\pi G \; \delta (t_1 - t_2)  \int_{\bm k} \frac{e^{i {\bm k}\cdot {\bm r}}}{{\bm k^2}} \delta_{ij}, \\
    \langle{\sigma_{ij}({\bm x_1},t_1)}{\sigma_{kl}({\bm x_2},t_2)}\rangle &= 32\pi G~P_{ij;kl} ~\delta(t_1-t_2)\int_{\bm k} \frac{e^{i {\bm k}\cdot {\bm r}}}{\bm k^2},
\end{align}
where $P_{ij;kl} \equiv \frac{1}{2} \left(\delta_{ik}\delta_{jl} + \delta_{il}\delta_{jk} - 2\delta_{ij}\delta_{kl}\right)$, ${\bm r} = {\bm x}_1 - {\bm x}_2$, which reflects the instantaneous nature of the potential-mode propagator \cite{Levi:2018nxp},
\begin{equation}
    \int \frac{dk_0}{2\pi}e^{-ik_0t}\int \frac{d^3{\bm k}}{(2\pi)^3}\frac{e^{i {\bm k}\cdot{\bm x}}}{{\bm k^2}}=\delta(t)\int_{\bm k}\frac{e^{i {\bm k}\cdot{\bm x}}}{{\bm k^2}}
\end{equation}
where we use the scaling of the orbital modes, $k_0 \sim v/r$ and $|{\bm k}| \sim 1/r$.
Beyond the leading approximation, subleading corrections arise from time-derivative interactions, which modify the instantaneous propagator.  

In the following, symmetry factors associated with identical fields are
accounted for at the level of the diagrams. We therefore omit explicit factors of $1/i!$ from vertices containing $i$ identical fields. The leading time-derivative correction to the $\phi$ propagator is then
\begin{align}
    & \begin{gathered}
     \begin{tikzpicture}
    \draw[black,  thick] (0,0) -- (2,0);
    \node [cross out,draw=black, thick] at (1.,0){};
    \end{tikzpicture}
    \end{gathered} = \frac{1}{4\pi G} \int d^4 x ~ \partial^{0}\phi\partial^{0}\phi
\end{align}
\subsection{Electromagnetic action}
\label{sec:Maxwell}
We consider the electromagnetic action in curved spacetime with a Feynman gauge fixing term:
\begin{flalign}
    \label{eq:sqaction}
	\mathcal{S}_{em} = \frac{-1}{16\pi} \int d^4 x \sqrt{-g} \, e^{-2a\varphi}
\left(  F^2  + 2(\partial_{\mu} A^\mu)^2 \right),
\end{flalign}
Here \(F_{\mu\nu} = \nabla_{\mu} A_{\nu} - \nabla_{\nu} A_{\mu}\), which reduces to \(F_{\mu\nu} = \partial_{\mu} A_{\nu} - \partial_{\nu} A_{\mu}\) due to the cancellation of the Christoffel symbols.
Substituting the KK decomposition~\eqref{eq:kka}, we expand the electromagnetic action perturbatively in the post-Newtonian regime.
From the quadratic terms in the expanded action, we obtain the propagators for the electromagnetic fields~\cite{Gupta:2022spq}:
\begin{align}
    \label{emscalarprop}
    \left< A_0 ({\bm x_1},t_1) A_0 ({\bm x_2},t_2)\right> = -4\pi  \; \delta(t_1 - t_2)  \int_{\bm k} \frac{e^{i {\bm k}\cdot {\bm r}}}{{\bm k^2}} ,\\
    \label{emvectorprop}
    \left< A_i ({\bm x_1},t_1) A_j({\bm x_2},t_2)\right> = 4 \pi \; \delta (t_1 - t_2)  \int_{\bm k} \frac{e^{i {\bm k}\cdot {\bm r}}}{{\bm k^2}} \delta_{ij},
\end{align}    
Subleading time-derivative terms generate corrections to the instantaneous propagators. The corresponding Feynman rules arise from time-derivative terms in the expanded action: 
\begin{align}
    \label{timederivative}
    & \begin{gathered}
     \begin{tikzpicture}
\draw[dashed, thick] (0,0) -- (2,0);
\node [cross out,draw=black, thick] at (1.,0){};
\end{tikzpicture}
    \end{gathered} =  \frac{-1}{4\pi} \int d^4 x ~ \partial_0 A_0\partial_0 A_0,\\
    & \begin{gathered}
     \begin{tikzpicture}
\draw[dashed, thick] (0,0) -- (2,0);
\draw[dashed, thick] (0,0.1) -- (2,0.1);
\node [cross out,draw=black, thick] at (1.,0.05){};
\end{tikzpicture}
    \end{gathered} =   \frac{1}{4\pi} \int d^4 x ~ \partial_{0}A_{i}\partial_{0}A_{j}\delta^{ij},
\end{align}

where the crossed vertices denote quadratic time-derivative insertions in the corresponding propagators. Additional electromagnetic interaction vertices without scalar field relevant at 2PN orders are given in Ref.~\cite{Gupta:2022spq}.

The coupling between the electromagnetic and dilaton fields arises from the factor $e^{-2a\varphi}$ in the electromagnetic action. Expanding this factor as and retaining the terms required through 2PN order in Eq~\eqref{eq:sqaction}, the corresponding interaction part of the action can be written as
\begin{align}
\label{eq:maxwellmetric}
\mathcal{S}_{em} &=  \frac{-1}{16\pi} \int d^4 x  \left(  -2\partial_{i}A_{0}\partial_{i}A_{0} + 2  (\partial_{i}A_{k}\partial_{i}A_{k})+2\partial_0 A_0\partial_0 A_0 \right. \nonumber \\ & \left. +4\phi\partial_{i}A_{0}\partial_{i}A_{0}\right) \left( - 2 a \varphi + 2 a^2 \varphi^2 \right)
\end{align}

We extract three-point self-interacting vertex and four-point vertex from Eq.~\eqref{eq:maxwellmetric},
\begin{align}
    \label{4pointvertex}
        & \begin{gathered}
     \begin{tikzpicture}
\draw[dashed,  thick] (0,0) -- (1.,1.);
\draw[dashed, thick] (0,0) -- (1.,-1.);
\draw[snake=zigzag,segment aspect=0,  thick] (-1,0) -- (0,0);
\filldraw[black] (0,0) circle (3pt) node[anchor=west]{};
\end{tikzpicture}
    \end{gathered} =  \frac{a}{4\pi} \int d^4 x \; \varphi \left( -2 (\partial^{i}A_{0}) (\partial_{i}A_{0})+ 2 (\partial_0 A_0)^2 \right)\\
    & \begin{gathered}
     \begin{tikzpicture}
\draw[dashed,  thick] (0,0) -- (1.,1.);
\draw[dashed,  thick] (0,-0.1) -- (1.1,1.);
\draw[dashed,  thick] (0,0) -- (1.,-1.);
\draw[dashed,  thick] (-0.1,0) -- (1.,-1.1);
\draw[snake=zigzag,segment aspect=0,  thick] (-1,0) -- (0,0);
\filldraw[black] (0,0) circle (3pt) node[anchor=west]{};
\end{tikzpicture}
    \end{gathered} =  \frac{a}{4\pi}  \int d^4x \; \varphi \; 2 \delta^{lk} (\partial^i A_k)(\partial_i A_l).\\
        & \begin{gathered}
     \begin{tikzpicture}
 \draw[snake=zigzag,segment aspect=0,  thick] (0,0) -- (1,1);
\draw[snake=zigzag,segment aspect=0,  thick] (0,0) -- (1,-1);
\draw[dashed,  thick] (-1.,1.) -- (0,0);
\draw[dashed,  thick] (-1.,-1.) -- (0,0);
\filldraw[black] (0,0) circle (3pt) node[anchor=west]{};
\end{tikzpicture}
    \end{gathered} =   \frac{a^2}{\pi} \int d^4x \; (\partial_{i}A_{0})(\partial^{i}A_{0}) \varphi ^2,\\
    \label{4pointvertex1}
    & \begin{gathered}
     \begin{tikzpicture}
\draw[black,  thick] (0,0) -- (1.,1.);
\draw[dashed,  thick] (-1.,1.) -- (0,0);
\draw[snake=zigzag,segment aspect=0,  thick] (0,0) -- (1,-1);
\draw[dashed,  thick] (-1.,-1.) -- (0,0);
\filldraw[black] (0,0) circle (3pt) node[anchor=west]{};
\end{tikzpicture}
    \end{gathered} =   \frac{a}{\pi} \int d^4x \; (\partial_{i}A_{0})(\partial^{i}A_{0}) \phi \; \varphi .
\end{align}
There are two kinds of two identical lines on the vertex, and we have dropped the factors $1/2!$ and $1/2!$. 
In what follows, these propagators and interaction vertices will be used to construct the Feynman diagrams contributing to the conservative two-body dynamics.

\subsection{Dilaton action}
\label{sec:Dilaton}
The scalar (dilaton) field is described by the action
\begin{flalign}
    \label{eq:sqaction}
	\mathcal{S}_{\varphi} = \frac{-1}{16\pi G} \int d^4 x \sqrt{-g} \left( 2g^{\mu\nu}\partial_\mu\varphi\partial_\nu\varphi \right),
\end{flalign}
Substituting the KK decomposition~\eqref{eq:kka}, we expand the action perturbatively in the post-Newtonian regime. The resulting expression contains kinetic terms as well as interactions between the dilaton and the NRG fields \(\phi\), \(\mathcal{A}_i\), and \(\sigma_{ij}\):
\begin{align}
S =&\int d^4x \, \frac{1}{16\pi G} \left(2\partial_0\varphi\partial_0\varphi -2\delta_{ij}\partial_i\varphi\partial_j\varphi -8 \phi \partial_0\varphi\partial_0\varphi \right. \nonumber \\
  & \left. -4 \mathcal{A}_i \partial_0\varphi\partial_i\varphi +2 (\sigma_{ij})\partial_i\varphi\partial_j\varphi  -\sigma_{l}^l \delta_{ij}\partial_i\varphi\partial_j\varphi \right)\nonumber \\
\end{align}
This action is in agreement with the result obtained independently using the EFTofPNG package~\cite{Levi:2017kzq}. From the quadratic part of the action, we obtain the propagator for the scalar field,
\begin{align}
    \label{dilatonscalar}
    \left< \varphi ({\bm x_1}, t_1) \varphi ({\bm x_2},t_2)\right> &= 4 \pi G \; \delta(t_1 - t_2)  \int_{\bm k}  \frac{e^{i {\bm k}\cdot {\bm r}}}{{\bm k^2}},
\end{align}

Subleading corrections arise from time-derivative interactions in the expanded action, which modify the instantaneous propagator. These terms correspond to quadratic self-interactions of the scalar field involving time derivatives.
\begin{align}
    \label{timederivative}
     \begin{gathered}
     \begin{tikzpicture}
     \draw[snake=zigzag,segment aspect=0,  thick] (0,0) -- (2,0);
\node [cross out,draw=black, very thick] at (1.,0){};
\end{tikzpicture}
    \end{gathered} =  \frac{1}{4\pi G} \int d^4 x ~ \partial^{0}\varphi\partial^{0}\varphi,
\end{align}

where the crosses represent the self-dilaton quadratic vertices, containing two time derivatives.\\
In addition, the expansion generates interaction vertices between the dilaton and the NRG fields. In particular, the dilaton couples to the scalar gravitational field \(\phi\), the gravito-magnetic field \(\mathcal{A}_i\), and the spatial metric perturbation \(\sigma_{ij}\) through derivative interactions. These vertices encode the nonlinear structure of the theory and will contribute to the conservative dynamics at higher PN orders.
\begin{align}
\label{fig:3pointv1}
    & \begin{gathered}
     \begin{tikzpicture}
\draw[black,  thick] (0,0) -- (1.,1.);
\draw[snake=zigzag,segment aspect=0,  thick] (0,0) -- (1,-1);
\draw[snake=zigzag,segment aspect=0,  thick] (-1,0) -- (0,0);
\filldraw[black] (0,0) circle (3pt) node[anchor=west]{};
\end{tikzpicture}
    \end{gathered} =  \frac{1}{8\pi G} \int d^4x \; (-8)\phi \partial_0\varphi\partial_0\varphi,\\
    \label{fig:3pointv2}
    & \begin{gathered}
     \begin{tikzpicture}
\draw[snake=zigzag,segment aspect=0,  thick] (0,0) -- (1,1);
\draw[snake=zigzag,segment aspect=0,  thick] (-1,0) -- (0,0);
\draw[snake=coil,segment aspect=0,  thick] (0,0) -- (1.,-1.);
\filldraw[black] (0,0) circle (3pt) node[anchor=west]{};
\end{tikzpicture}
    \end{gathered} =  \frac{1}{8\pi G}  \int d^4x \; (-4) \mathcal{A}_i \partial_0\varphi\partial_i\varphi,\\
    \label{fig:3pointv5}
    & \begin{gathered}
     \begin{tikzpicture}
\draw[snake=coil,segment aspect=0,  thick] (0,0) -- (1.,1.);
\draw[snake=coil,segment aspect=0,  thick] (0,-0.1) -- (1.1,1.);
\draw[snake=zigzag,segment aspect=0,  thick] (0,0) -- (1,-1);
\draw[snake=zigzag,segment aspect=0,  thick] (-1,0) -- (0,0);
\filldraw[black] (0,0) circle (3pt) node[anchor=west]{};
\end{tikzpicture}
    \end{gathered} =   \frac{1}{8\pi G}  \int d^4x \; 2 (\sigma_{ij})\partial_i\varphi\partial_j\varphi  -\sigma_{l}^l \delta_{ij}\partial_i\varphi\partial_j\varphi.
\end{align}

\begin{table}
\centering
\caption{This table shows the order of the terms that contribute at given PN orders.}
\label{fig:table1}
\begin{ruledtabular}
\begin{tabular}{@{\hspace{0.6cm}}l c r@{\hspace{0.6cm}}}
0PN & 1PN & 2PN  \\[2pt]
\hline
$G$ & $G^2$ & $G^3$ \\
 & $G v^2 $ & $G v^4$ \\
 &  & $G^2 v^2$ \\
$q^2$ & $q^2v^2$ & $q^2v^4$ \\
&  & $q^4v^2$ \\
 & $Gq^2$ & $Gq^2v^2$  \\
&  & $G^2q^2$  \\
& & $Gq^4$  \\
\end{tabular}
\end{ruledtabular}
\end{table}

\subsection{Power counting and Feynman diagrams}
\label{subsection:powercounting}

For bound binary systems, the virial relation provides the basic power-counting rule for the PN expansion,
\begin{equation}
    v^2 \sim \frac{Gm}{r} \sim \frac{q^2}{mr},
\end{equation}
or, equivalently,
\begin{equation}
    mv^2 \sim \frac{Gm^2}{r} \sim \frac{q^2}{r}.
\end{equation}
We independently track powers of $G$, $q$, and $v$ to determine the PN order of each contribution.

The dilaton field $\varphi$ is counted in the same way as the NRG scalar gravitational field $\phi$. In particular, its leading worldline coupling carries no additional powers of the orbital velocity, and dilaton exchange follows the same $G$ power counting as the corresponding gravitational scalar exchange. We therefore use powers of $G$ to organize diagrams involving both gravitational and dilaton fields, while the electromagnetic interactions are additionally classified by their explicit powers of the charges $q_A$. The dimensionless scalar-response parameters $\alpha_A$, $\beta_A$, and $\beta'_A$, as well as the dilaton coupling $a$, are taken to be of $\mathcal O(v^0)$.

At leading order, scalar-mediated interactions from dilaton exchange contribute at $\mathcal{O}(G)$. As summarized in Table~\ref{fig:table1}, these constitute the leading long-range interactions beyond pure gravity. At 1PN order, the relevant contributions scale as $\mathcal{O}(G^2)$, $\mathcal{O}(Gv^2)$, and $\mathcal{O}(Gq^2)$. At 2PN order, we include all diagrams scaling as $\mathcal{O}(Gv^4)$, $\mathcal{O}(G^2v^2)$, $\mathcal{O}(Gq^2v^2)$, $\mathcal{O}(G^3)$, $\mathcal{O}(G^2q^2)$, and $\mathcal{O}(Gq^4)$. The purely gravitational contributions have been computed previously in Ref.~\cite{Gilmore:2008gq}, while the corresponding Einstein-Maxwell contributions were obtained in Ref.~\cite{Gupta:2022spq}.

We construct all relevant Feynman-diagram topologies through $\mathcal O(G^3)$ following standard power-counting rules. A worldline vertex involving $n$ gravitational fields contributes a factor $G^{n/2}$, while a bulk $n$-graviton vertex scales as $G^{n/2-1}$~\cite{Gilmore:2008gq}. Once the dependence on $G$ is established, the corresponding velocity scaling is determined from the fields and interaction vertices entering each diagram.

Velocity power counting is performed using the NRG fields $\phi$, $\mathcal A_i$, and $\sigma_{ij}$, whose leading worldline couplings scale as $\mathcal O(v^0)$, $\mathcal O(v)$, and $\mathcal O(v^2)$, respectively. The dilaton $\varphi$ follows the same leading counting as $\phi$. For the electromagnetic sector, the temporal and spatial components of the gauge field scale as $A_0\sim\mathcal O(v^0)$ and $A_i\sim\mathcal O(v)$, respectively. Additional velocity suppression arises from time derivatives acting on propagators and interaction vertices. Since $\partial_t\sim v\,\partial_i$, a pair of time derivatives introduces an additional factor of $\mathcal O(v^2)$.

Using these rules, we systematically generate the relevant diagram topologies and populate them with the NRG fields $\phi$, $\mathcal A_i$, and $\sigma_{ij}$, the electromagnetic fields $A_0$ and $A_i$, and the dilaton field $\varphi$ in all allowed combinations. The overall scaling of each diagram is then determined by its worldline and bulk vertices, propagators, charge insertions, and time derivatives.

The resulting diagrams are organized as follows: the leading-order Coulomb and scalar-exchange contributions are shown in Fig.~\ref{fig:coulombfig}, the 1PN corrections in Fig.~\ref{fig:fig1}, and the 2PN contributions in Figs.~\ref{fig:Gv^4}, \ref{fig:G^2v^2}, \ref{fig:Gq^2v^2}, \ref{fig:G^3}, \ref{fig:G^2q^2} and \ref{fig:Gq^4}. Diagrams containing closed gravitational loops are omitted, as they correspond to quantum corrections \cite{Holstein:2004dn}.

\section{0PN Potential}
\label{section:0PN}
\begin{figure}[hbt!]
\centering
\includegraphics{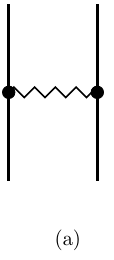}
\caption{Leading-order single-dilaton-exchange diagram.}
\label{fig:coulombfig}
\end{figure}

In this section, we compute the leading-order scalar-mediated potential arising from dilaton exchange.
The corresponding contribution is generated by the single-dilaton-exchange diagram shown in Fig.~\ref{fig:coulombfig}. Using the worldline coupling to $\varphi$ from Sec.~\ref{sec:DilatonParticle} together with the corresponding instantaneous propagator in Eq.~\eqref{dilatonscalar} and the Fourier identity in Eq.~\eqref{eq:ft}, we obtain
\begin{align}
  \mathrm{Diagram} \; \ref{fig:coulombfig} \mathrm{(a)} = \frac{G m_1 m_2 \alpha_1 \alpha_2}{r}.
\end{align}
Adding the leading-order contributions from the gravitational and electromagnetic sectors, the total 0PN Lagrangian in EMd theory reads
\begin{equation}
  \mathcal{L}_{0\text{PN}}
    = \frac{G_{12} m_1 m_2}{r} - \frac{q_1 q_2}{r},
\end{equation}
where $ G_{12} \equiv G\left(1 + \alpha_1 \alpha_2\right)$ defines the effective gravitational constant used throughout.

\section{1PN Potential}
\label{sec:1PN}
\begin{figure}[hbt!]
\centering
   \includegraphics{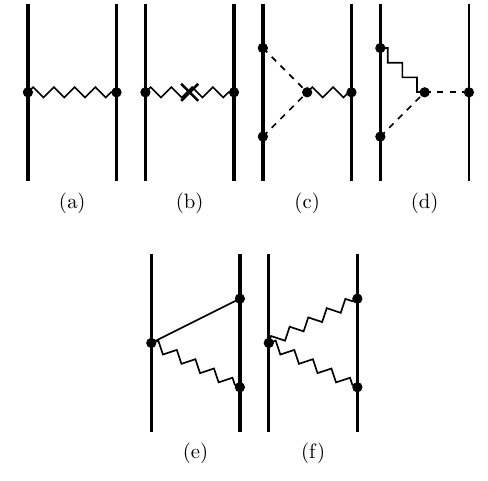} 
\caption{At 1PN order, diagrams (a) and (b) contribute at $\mathcal{O}(Gv^2)$, diagrams (c) and (d) at $\mathcal{O}(Gq^2)$, and diagrams (e) and (f) at $\mathcal{O}(G^2)$.}
\label{fig:fig1}  
\end{figure}

In this section, we compute the contributions to the conservative dynamics at 1PN order. The relevant diagrams fall into three categories: $\mathcal O(Gv^2)$, $\mathcal O(Gq^2)$, and $\mathcal O(G^2)$, as shown in Fig.~\ref{fig:fig1}. Their evaluation involves the worldline and bulk interaction vertices, together with the corresponding propagators, time-derivative corrections, and loop and Fourier integrals. The symmetry factors of the diagrams are determined as described in \cite{Gilmore:2008gq}; alternatively, they can be computed using Wick’s theorem without considering time ordering. We specify the symmetry factor for a diagram when it is not equal to 1. 

\subsection{$Gv^2$ order diagrams}
We begin with diagram (a) in Fig.~\ref{fig:fig1}, which arises from velocity-suppressed worldline couplings. Its contribution is

\begin{flalign}
\begin{split}
\mathrm{Diagram} &\; \ref{fig:fig1} \mathrm{(a)} =   -\int dt \frac{1}{2}\frac{G m_1 m_2}{r}\alpha_1\alpha_2\left({\bm v}_1 \cdot {\bm v}_1 + {\bm v}_2 \cdot {\bm v}_2 \right) \\
\end{split}
\end{flalign}

Diagram~(b) in Fig.~\ref{fig:fig1} arises from the time-derivative
correction to the dilaton propagator. Using the Feynman rules of
Sec.~\ref{sec:EFT FRAMEWORK} and the identities collected in Appendix~\ref{app:useful-formulas},
its contribution is

\begin{align}
    \mathrm{Diagram \; } \ref{fig:fig1} \mathrm{(b)} &=  \int d t G m_1 m_2 \frac{\alpha_1 \alpha_2}{2r} \left( {\bm v}_1 \cdot {\bm v}_2 - ({\bm v}_1 \cdot {\bm n})({\bm v}_2 \cdot {\bm n}) \right).
\end{align}

\subsection{$Gq^2$ order diagrams}

We illustrate the evaluation of the $\mathcal O(Gq^2)$ contributions using diagram~(c) in Fig.~\ref{fig:fig1}. Applying the relevant worldline couplings, propagators, and three-point bulk vertex, and performing the integration over the bulk interaction point $d^3x$ and time, we obtain
\begin{flalign}
\begin{split}
	\mathrm{Diagram \; } \ref{fig:fig1} \mathrm{(c)} 
    &= \mathcal C  \int dt  \int_{k_2}  \int_{k_3}  \; \frac{e^{i {\bm k_3}\cdot ({{\bm x_2}}-{{\bm x_1}})}}{({\bm k_2}+{\bm k_3})^2}\; \frac{(k_2^i+k_3^i)k_2^j}{{\bm k_2}^2{\bm k_3}^2}\;\delta_{ij}.\\
\end{split}
\end{flalign}
where $ \mathcal C = (4\pi)^2 Gm_2q_1^2 a \alpha_2$. Using the one-loop tensor master integral, Eq.~\eqref{eq:1loopvec}, together with the tensor Fourier identity, Eq.~\eqref{tensorfourierindentity}, gives
\begin{flalign}
\begin{split}
	\mathrm{Diagram \; } \ref{fig:fig1} \mathrm{(c)} 
    &=     -\int dt \; \frac{G a m_2q_1^2 \alpha_2}{r^2}.
\end{split}
\end{flalign}
Including the symmetry factor $1/2$, the contribution becomes
\begin{equation}
    \mathrm{Diagram \; } \ref{fig:fig1} \mathrm{(c)} =     -\int dt \; \frac{G a m_2q_1^2 \alpha_2}{2r^2}.
\end{equation}
We compute the diagram \ref{fig:fig1}(d) similarly,
\begin{flalign}
\begin{split}
	\mathrm{Diagram \; } \ref{fig:fig1} \mathrm{(d)} 
	    = \int dt \frac{G a m_2 q_1q_2 \alpha_2}{ r^2}.
\end{split}
\end{flalign}

\subsection{$G^2$ order diagrams}
The remaining two diagrams contribute at $\mathcal O(G^2)$. Using the corresponding worldline and bulk vertices and the instantaneous propagators, diagram~\ref{fig:fig1}(e) gives
\begin{flalign}
\begin{split}
	\mathrm{Diagram \; } \ref{fig:fig1} \mathrm{(e)} 
    = -\int dt \frac{G^2 m_1 m_2^2 \alpha_1 \alpha_2}{ r^2}.
\end{split}
\end{flalign}

We compute the diagram \ref{fig:fig1}(f) similarly which has symmetry factor of $\frac{1}{2}$,
\begin{flalign}
\begin{split}
	\mathrm{Diagram \; } \ref{fig:fig1} \mathrm{(f)} 
    = -\int dt \frac{G^2 m_1 m_2^2}{2 r^2}\left(\alpha_1^2 \alpha_2^2 + \alpha_2^2 \beta_1  \right).
\end{split}
\end{flalign}

Combining the contributions from the diagrams above with the known gravitational and electromagnetic contributions, and including the terms obtained under $1\leftrightarrow2$ together with $\bm n\rightarrow-\bm n$, we obtain the complete conservative Lagrangian at 1PN order:
\begin{widetext}
\begin{align}
\mathcal{L}_{\mathrm{1PN}}
&= \frac{1}{8}m_1\bm v_1^4
 + \frac{q_1q_2}{4r} \Big[   \bm v_1\!\cdot\!\bm v_2 +(\bm n\!\cdot\!\bm v_1)(\bm n\!\cdot\!\bm v_2)\Big]
\nonumber\\
&\quad
 + \frac{G_{12}m_1m_2}{r}
 \Big[ \frac{3}{2}\bm v_1^2 -\frac{7}{4}\bm v_1\!\cdot\!\bm v_2   -\frac{1}{4}(\bm n\!\cdot\!\bm v_1)     (\bm n\!\cdot\!\bm v_2)+\bar\gamma_{12} \big(\bm v_1^2-\bm v_1\!\cdot\!\bm v_2\big)  \Big]
\nonumber\\
&\quad + \frac{G\hat m_1}{r^2}  \left(q_1q_2-\frac{1}{2}q_2^2\right) - \frac{G_{12}^2m_1^2m_2}{2r^2} \left(1+2\bar\beta_2\right)+ (1\leftrightarrow2).
\label{eq:L1PN}
\end{align}
\end{widetext}
Here, $(1\leftrightarrow2)$ denotes the contribution obtained by interchanging the particle labels $1\leftrightarrow2$, together with $\bm n\rightarrow-\bm n$. We have also introduced
\begin{align}
\bar\gamma_{12}
&= -\frac{2\alpha_1\alpha_2}{1+\alpha_1\alpha_2},\\
\bar\beta_1
&= \frac{1}{2}
\frac{\beta_1\alpha_2^2}{(1+\alpha_1\alpha_2)^2},
\end{align}
with $\bar\beta_2$ obtained by $1\leftrightarrow2$. 
We further define the dressed mass
\begin{equation}
\hat m_A \equiv m_A\left(1+a\alpha_A\right).
\end{equation}

The 1PN Lagrangian in Eq.~\eqref{eq:L1PN} agrees with the known 1PN dynamics of EMd binaries~\cite{Julie:2017rpw,Khalil:2018aaj}. In the electrically neutral limit, $q_A=0$, it reduces to the corresponding 1PN scalar-tensor Lagrangian~\cite{Almeida:2024uph}. Setting the scalar couplings to zero, $\alpha_A=\beta_A=0$, together with $a=0$, recovers
the Einstein-Maxwell result~\cite{Patil:2020dme,Gupta:2022spq}, while setting both the electric and scalar couplings to zero reproduces the 1PN conservative dynamics of general relativity~\cite{Gilmore:2008gq}.

\section{2PN order}
\label{sec:2PN}
In this section, we compute the contributions to the conservative dynamics at 2PN order. The relevant diagrams fall into six categories: $\mathcal{O}(Gv^4)$, $\mathcal{O}(G^2v^2)$, $\mathcal{O}(Gq^2v^2)$, $\mathcal{O}(G^3)$, $\mathcal{O}(G^2q^2)$, and $\mathcal{O}(Gq^4)$. The evaluation of these diagrams closely follows the techniques used in our previous Einstein-Maxwell calculation~\cite{Gupta:2022spq}. These contributions are presented in the subsections below.

The $\mathcal{O}(Gv^4)$ diagrams include higher-order corrections to the diagrams in Fig.~\ref{fig:coulombfig}, together with contributions involving time-derivative insertions, which are evaluated similarly to diagram 2(b) in Fig.~\ref{fig:fig1}. The $\mathcal{O}(Gq^2v^2)$ and $\mathcal{O}(G^2v^2)$ diagrams involve one-loop integrals and time-derivative insertions, analogous to the diagrams at 1PN order. The $\mathcal{O}(G^3)$, $\mathcal{O}(Gq^4)$, and $\mathcal{O}(G^2q^2)$ diagrams involve two-loop integrals, which can be categorized into three types. The first type consists of factorizable two-loop integrals that decompose into products of two one-loop integrals, allowing each one-loop integral to be computed separately. These integrals are relatively straightforward to evaluate. The second type consists of nested two-loop integrals, in which one loop is nested within another, requiring successive evaluation: first the inner loop and then the outer loop. The third type consists of irreducible two-loop integrals, which can be reduced, using integration-by-parts identities~\cite{Smirnov:2004ym}, to sums of factorizable and nested two-loop integrals. We explicitly indicate the symmetry factor whenever it differs from unity, and all diagrammatic results presented below include the corresponding symmetry factors.

\begin{figure}[h]
    \centering
\includegraphics{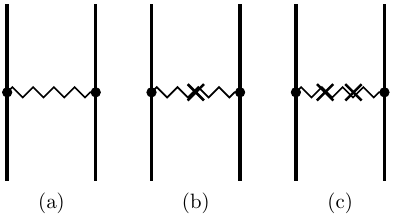} 
\caption{Diagrams contributing at $\mathcal{O}(Gv^4)$, including up to two time-derivative insertions.}
\label{fig:Gv^4}
\end{figure}

\subsection{$Gv^4$ order diagrams}
The three diagrams contributing at $\mathcal{O}(Gv^4)$ are shown in Fig.~\ref{fig:Gv^4}. Diagram \ref{fig:Gv^4}(a) arises from the higher-order expansion of the worldline couplings, whereas diagrams \ref{fig:Gv^4}(b) and \ref{fig:Gv^4}(c) contain one and two time-derivative insertions, respectively. Their contributions are :
\begin{widetext}
\begin{align}
\mathrm{Diagram \;} \ref{fig:Gv^4} \mathrm{(a)}  &=\int d t\; G \alpha_1 \alpha_2 m_1 m_2  \frac{ \left(-\frac{1}{8} \left({\bm v}_1 \cdot {\bm v}_1\right){}^2+\frac{1}{4} \left({\bm v}_1 \cdot {\bm v}_1\right) \left({\bm v}_2 \cdot {\bm v}_2\right)-\frac{1}{8} \left({\bm v}_2 \cdot {\bm v}_2\right){}^2\right)}{r}
\end{align}

\begin{align}
\mathrm{Diagram \;} \ref{fig:Gv^4} \mathrm{(b)}  & = \int d t\;G \alpha_1 \alpha_2 m_1 m_2  \left(-\frac{1}{2} ({\bm v}_1 \cdot {\bm a}_1) ({\bm v}_2 \cdot {\bm n})+\frac{1}{2} ({\bm v}_2 \cdot {\bm a}_2) ({\bm v}_1 \cdot {\bm n})+\frac{({\bm v}_1 \cdot {\bm v}_1) ({\bm v}_1 \cdot {\bm n}) ({\bm v}_2 \cdot {\bm n})}{4 r} \right. \nonumber \\
&\left. +\frac{({\bm v}_2 \cdot {\bm v}_2) ({\bm v}_1 \cdot {\bm n}) ({\bm v}_2 \cdot {\bm n})}{4 r} -\frac{({\bm v}_1 \cdot {\bm v}_1) ({\bm v}_1 \cdot {\bm v}_2)}{4 r}-\frac{({\bm v}_2 \cdot {\bm v}_2) ({\bm v}_1 \cdot {\bm v}_2)}{4 r}\right)
\end{align}

\begin{align}
 \mathrm{Diagram \; } \ref{fig:Gv^4} \mathrm{(c)} & = \int d t\; G \alpha_1 \alpha_2 m_1 m_2  \left(-\frac{1}{8} r ({\bm a}_1 \cdot {\bm n}) ({\bm a}_2 \cdot {\bm n})+\frac{1}{8} ({\bm a}_2 \cdot {\bm n}) \left({\bm v}_1 \cdot {\bm n}\right){}^2-\frac{1}{4} ({\bm v}_1 \cdot {\bm a}_2) ({\bm v}_1 \cdot {\bm n})-\frac{1}{8} ({\bm a}_1 \cdot {\bm n}) \left({\bm v}_2 \cdot {\bm n}\right){}^2 \right. \nonumber \\
&\left. -\frac{1}{8} ({\bm v}_1 \cdot {\bm v}_1) ({\bm a}_2 \cdot {\bm n}) + \frac{1}{4} ({\bm a}_1 \cdot {\bm v}_2) ({\bm v}_2 \cdot {\bm n})+\frac{1}{8} ({\bm v}_2 \cdot {\bm v}_2) ({\bm a}_1 \cdot {\bm n})-\frac{1}{8} ({\bm a}_1 \cdot {\bm a}_2) r+\frac{3 \left({\bm v}_2 \cdot {\bm n}\right){}^2 \left({\bm v}_1 \cdot {\bm n}\right){}^2}{8 r}\right. \nonumber \\
&\left. -\frac{({\bm v}_2 \cdot {\bm v}_2) \left({\bm v}_1 \cdot {\bm n}\right){}^2}{8 r}  -\frac{({\bm v}_1 \cdot {\bm v}_2) ({\bm v}_2 \cdot {\bm n}) ({\bm v}_1 \cdot {\bm n})}{2 r}-\frac{({\bm v}_1 \cdot {\bm v}_1) \left({\bm v}_2 \cdot {\bm n}\right){}^2}{8 r}+\frac{\left({\bm v}_1 \cdot {\bm v}_2\right){}^2}{4 r}+\frac{({\bm v}_1 \cdot {\bm v}_1) ({\bm v}_2 \cdot {\bm v}_2)}{8 r}\right)
\end{align}
\end{widetext}

\subsection{$G^2v^2$ order diagrams}
\begin{figure*}[htbp]
\centering
\includegraphics{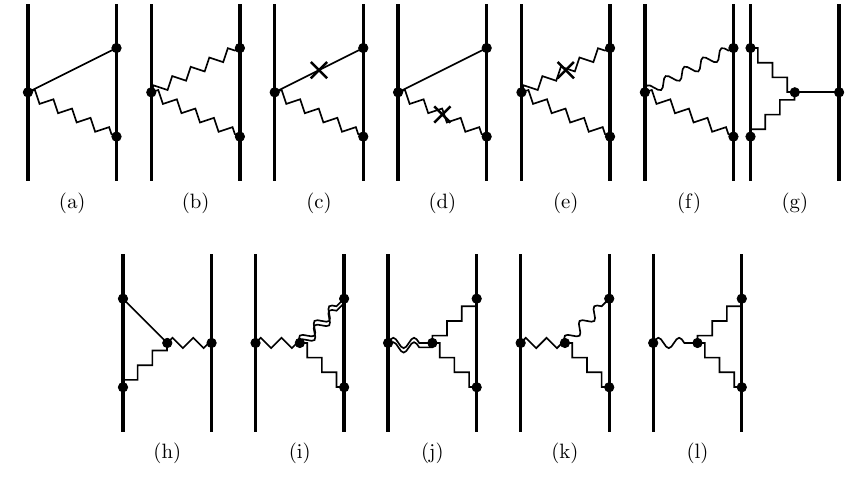}
\caption{The 12 diagrams contributing at $\mathcal{O}(G^2v^2)$. Diagrams (b), (g), (j), and (l) have a symmetry factor of $1/2$, while the remaining diagrams have a symmetry factor of unity.}
\label{fig:G^2v^2}
\end{figure*}

At $\mathcal{O}(G^2v^2)$, there are 12 diagrams, as shown in Fig.~\ref{fig:G^2v^2}. These diagrams involve different combinations of field exchanges, including one-loop contributions and diagrams with time-derivative insertions. Their evaluation follows the techniques described at 1PN order. The resulting contributions are
\begin{widetext}
\begin{flalign}
\begin{split}
	\mathrm{Diagram \; } \ref{fig:G^2v^2} \mathrm{(a)}
	&= -\int dt\, \frac{G^{2}\,m_1\,m_2^{2}\,\alpha_1\alpha_2}{r^{2}}
\left[\frac{3}{2}(\bm v_1\cdot\bm v_1)+(\bm v_2\cdot\bm v_2)\right]
\end{split}
\end{flalign}

\begin{flalign}
\begin{split}
	\mathrm{Diagram \; } \ref{fig:G^2v^2} \mathrm{(b)}
	&= \int dt\, \frac{G^{2}\,m_1\,m_2^{2}}{r^{2}}
\left[\frac{1}{4}(\bm v_1\cdot\bm v_1)\,\alpha_1^{2}\alpha_2^{2} +\frac{1}{2}(\bm v_2\cdot\bm v_2)\,\alpha_1^{2}\alpha_2^{2}
+\frac{1}{4}(\bm v_1\cdot\bm v_1)\,\alpha_2^{2}\beta_1+\frac{1}{2}(\bm v_2\cdot\bm v_2)\,\alpha_2^{2}\beta_1\right]
\end{split}
\end{flalign}
\end{widetext}

\begin{widetext}
\begin{flalign}
\begin{split}
\mathrm{Diagram}\;\ref{fig:G^2v^2}\mathrm{(c)}
&= \int dt\,\frac{G^2m_1m_2^2\alpha_1\alpha_2}{r^2}\left[-\frac{1}{2}\bm v_1\!\cdot\!\bm v_2+(\bm v_1\!\cdot\!\bm n)(\bm v_2\!\cdot\!\bm n)-\frac{1}{2}(\bm v_2\!\cdot\!\bm n)^2\right]
\end{split}
\end{flalign}

\begin{flalign}
\begin{split}
\mathrm{Diagram}\;\ref{fig:G^2v^2}\mathrm{(d)}
&= \int dt\,\frac{G^2m_1m_2^2\alpha_1\alpha_2}{r^2}\left[-\frac{1}{2}\bm v_1\!\cdot\!\bm v_2+(\bm v_1\!\cdot\!\bm n)(\bm v_2\!\cdot\!\bm n)-\frac{1}{2}(\bm v_2\!\cdot\!\bm n)^2\right]
\end{split}
\end{flalign}

\begin{flalign}
\begin{split}
	\mathrm{Diagram \; } \ref{fig:G^2v^2} \mathrm{(e)}
	&= \int dt\, \frac{G^{2}\,m_1\,m_2^{2}}{r^{2}}\left[-\frac{1}{2}(\bm v_1\cdot\bm v_2)\alpha_1^{2}\alpha_2^{2}-\frac{1}{2}(\bm v_1\cdot\bm v_2)\alpha_2^{2}\beta_1+(\bm v_1\cdot\bm n)(\bm v_2\cdot\bm n)\alpha_1^{2}\alpha_2^{2}-\frac{1}{2}(\bm v_2\cdot\bm n)^2\alpha_1^{2}\alpha_2^{2} \right. \nonumber \\ & \left. +(\bm v_1\cdot\bm n)(\bm v_2\cdot\bm n)\alpha_2^{2}\beta_1-\frac{1}{2}(\bm v_2\cdot\bm n)^2\alpha_2^{2}\beta_1\right]
\end{split}
\end{flalign}

\begin{flalign}
\begin{split}
	\mathrm{Diagram \; } \ref{fig:G^2v^2} \mathrm{(f)}
&=\int dt\, \frac{4\,G^{2}\,m_1\,m_2^{2}\,\alpha_1\alpha_2}{r^{2}}(\bm v_1\cdot\bm v_2)
\end{split}
\end{flalign}

\begin{flalign}
\begin{split}
	\mathrm{Diagram \; } \ref{fig:G^2v^2} \mathrm{(g)}
	&= \int dt\, \frac{G^{2}\,m_1\,m_2^{2}\,\alpha_2^{2}}{2\,r^{2}}\left[-(\bm v_2\cdot\bm v_2)+(\bm v_2\cdot\bm n)^2\right]
	\end{split}
\end{flalign}

\begin{flalign}
\begin{split}
	\mathrm{Diagram \; } \ref{fig:G^2v^2} \mathrm{(h)}
	&= \int dt\, \frac{G^{2}\,m_1^{2}\,m_2\,\alpha_1\alpha_2}{r^{2}}\left[2(\bm v_1\cdot\bm v_2)-4(\bm v_1\cdot\bm n)(\bm v_2\cdot\bm n)\right]
\end{split}
\end{flalign}

\begin{flalign}
\begin{split}
	\mathrm{Diagram \; } \ref{fig:G^2v^2} \mathrm{(i)}
&= \int dt\, \frac{2\,G^{2}\,m_1^{2}\,m_2\,\alpha_1\alpha_2}{r^{2}}\left[(\bm v_1\cdot\bm v_1)-2(\bm v_1\cdot\bm n)^2\right]
\end{split}
\end{flalign}

\begin{flalign}
\begin{split}
	\mathrm{Diagram \; } \ref{fig:G^2v^2} \mathrm{(j)}
&= \int dt\, \frac{G^{2}\,m_1\,m_2^{2}\,\alpha_2^{2}}{2\,r^{2}}\left[-(\bm v_1\cdot\bm v_1)+(\bm v_1\cdot\bm n)^2\right]
\end{split}
\end{flalign}

\begin{flalign}
\begin{split}
	\mathrm{Diagram \; } \ref{fig:G^2v^2} \mathrm{(k)}
&= \int dt\, \frac{G^{2}\,m_1^{2}\,m_2\,\alpha_1\alpha_2}{r^{2}}\left[-2(\bm v_1\cdot\bm v_1)-2(\bm v_1\cdot\bm v_2)+4(\bm v_1\cdot\bm n)^2+4(\bm v_1\cdot\bm n)(\bm v_2\cdot\bm n)\right]
\end{split}
\end{flalign}

\begin{flalign}
\begin{split}
	\mathrm{Diagram \; } \ref{fig:G^2v^2} \mathrm{(l)}
&= \int dt\, \frac{G^{2}\,m_1\,m_2^{2}\,\alpha_2^{2}}{r^{2}}\left[(\bm v_1\cdot\bm v_2)-(\bm v_1\cdot\bm n)(\bm v_2\cdot\bm n)\right]
\end{split}
\end{flalign}
\end{widetext}

\subsection{$Gq^2v^2$ order diagrams}
\begin{figure*}[htbp]
\centering
\includegraphics{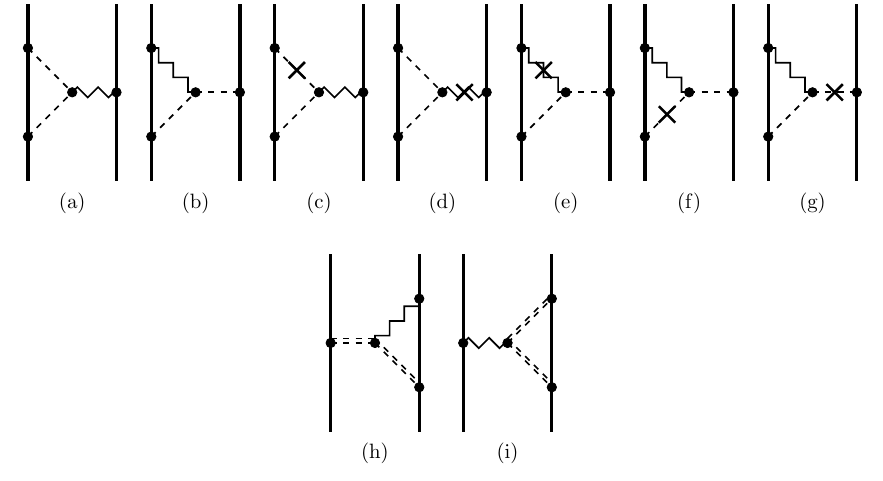}
\caption{The nine diagrams contributing at $\mathcal{O}(Gq^2v^2)$. Diagrams (a), (d), and (i) have a symmetry factor of $1/2$, while the remaining diagrams have a symmetry factor of unity.}
\label{fig:Gq^2v^2}
\end{figure*}

At $\mathcal{O}(Gq^2v^2)$, there are nine diagrams, as shown in Fig.~\ref{fig:Gq^2v^2}. These diagrams involve different combinations of gravitational, electromagnetic, and scalar interactions, including one-loop contributions and diagrams with time-derivative insertions. Their evaluation follows the techniques described at 1PN order. The resulting contributions are

\begin{widetext}
\begin{flalign}
\begin{split}
	\mathrm{Diagram \; } \ref{fig:Gq^2v^2} \mathrm{(a)}
	&= \int dt\, \frac{a\,G\,m_2\,q_1^{2}\,\alpha_2}{4\,r^{2}} \left[ (\bm v_1 \cdot \bm v_1) +(\bm v_2 \cdot \bm v_2) -(\bm v_1 \cdot \bm n)^2 \right].
\end{split}
\end{flalign}
\begin{flalign}
\begin{split}
	\mathrm{Diagram \; } \ref{fig:Gq^2v^2} \mathrm{(b)}
	&= \int dt\, \frac{a\,G\,m_1\,q_1 q_2\,\alpha_1}{r^{2}}\left[2(\bm v_1\cdot\bm n)(\bm v_2\cdot\bm n) -\frac{1}{2}(\bm v_1\cdot\bm v_1)-(\bm v_1\cdot\bm v_2)\right]
\end{split}
\end{flalign}

\end{widetext}

\begin{flalign}
\begin{split}
	\mathrm{Diagram \; } \ref{fig:Gq^2v^2} \mathrm{(c)}
	&= -\int dt\, \frac{a\,G\,m_2\,q_1^{2}\,\alpha_2}{4\,r^{2}}\left[(\bm v_1\cdot\bm v_1)+(\bm v_1\cdot\bm n)^2\right],
\end{split}
\end{flalign}

\begin{flalign}
\begin{split}
	\mathrm{Diagram \; } \ref{fig:Gq^2v^2} \mathrm{(d)}
	&= \int dt\, \frac{a\,G\,m_2\,q_1^{2}\,\alpha_2}{r^{2}}\left[-\frac{1}{2}(\bm v_1\cdot\bm v_2)+(\bm v_1\cdot\bm n)(\bm v_2\cdot\bm n)\right],
\end{split}
\end{flalign}

\begin{flalign}
\begin{split}
	\mathrm{Diagram \; } \ref{fig:Gq^2v^2} \mathrm{(e)}
	&= \int dt\, \frac{a\,G\,m_1\,q_1 q_2\,\alpha_1}{4\,r^{2}}\left[(\bm v_1\cdot\bm v_1)+(\bm v_1\cdot\bm n)^2\right]
\end{split}
\end{flalign}
\begin{widetext}
\begin{flalign}
\begin{split}
	\mathrm{Diagram \; } \ref{fig:Gq^2v^2} \mathrm{(f)}
	&= \int dt\, a\,G\,m_1 q_1 q_2\,\alpha_1\left[-\frac{(\bm v_1\cdot\bm v_1)}{4r^2}+\frac{(\bm v_1\cdot\bm v_2)}{r^2}+\frac{3(\bm v_1\cdot\bm n)^2}{4r^2}-\frac{2(\bm v_1\cdot\bm n)(\bm v_2\cdot\bm n)}{r^2}\right]
\end{split}
\end{flalign}

\begin{flalign}
\begin{split}
	\mathrm{Diagram \; } \ref{fig:Gq^2v^2} \mathrm{(g)}
	&= \int dt\, \frac{a\,G\,m_1\,q_1 q_2\,\alpha_1}{r^{2}}\left[(\bm v_1\cdot\bm v_2)-2(\bm v_1\cdot\bm n)(\bm v_2\cdot\bm n)\right]
\end{split}
\end{flalign}
\end{widetext}

\begin{flalign}
\begin{split}
	\mathrm{Diagram \; } \ref{fig:Gq^2v^2} \mathrm{(h)}
	&= -\int dt\,\frac{a\,Gm_2q_1q_2\alpha_2}{r^2}\left(\mathbf{v}_1\!\cdot\!\mathbf{v}_2\right)
\end{split}
\end{flalign}

\begin{flalign}
\begin{split}
	\mathrm{Diagram \; } \ref{fig:Gq^2v^2} \mathrm{(i)}
	&= \int dt\, \frac{a\,G\,m_1\,q_2^{2}\,\alpha_1}{2\,r^{2}}
(\bm v_2\cdot\bm v_2)
\end{split}
\end{flalign}



\subsection{$G^3$ order diagrams}
\begin{figure*}[htbp]
\centering
\includegraphics{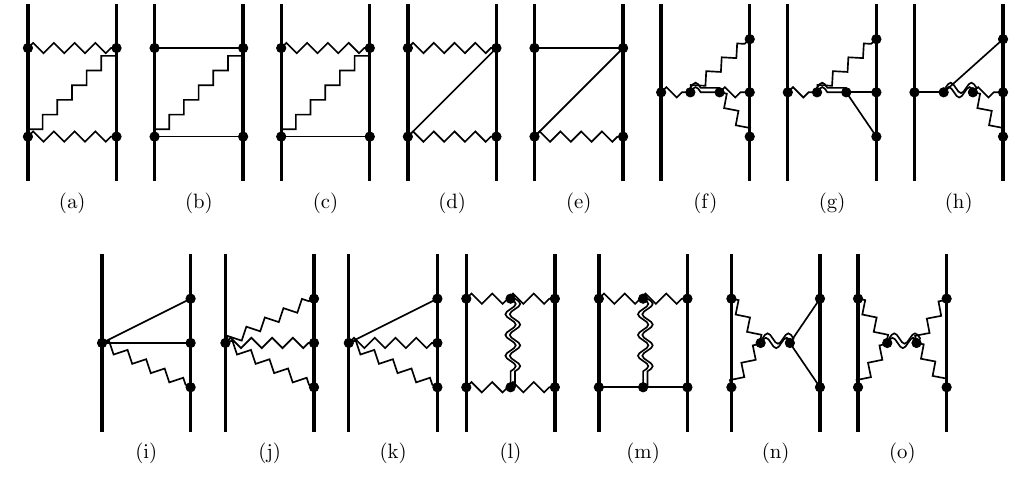}
\caption{The 15 diagrams contributing at $\mathcal{O}(G^3)$. Diagrams (f), (g), (h), (i), (k), and (l) have a symmetry factor of $1/2$, diagrams (n) and (o) have a symmetry factor of $1/4$, and diagram (j) has a symmetry factor of $1/6$, while the remaining diagrams have a symmetry factor of unity.}
\label{fig:G^3}
\end{figure*}

At $\mathcal{O}(G^3)$, there are 15 two-loop diagrams, as shown in Fig.~\ref{fig:G^3}. They can be classified into factorizable, nested, and irreducible two-loop topologies and are evaluated using the techniques used in our previous Einstein-Maxwell calculation~\cite{Gupta:2022spq}, including the reduction of irreducible two-loop integrals using integration-by-parts identities. The resulting contributions are

\begin{flalign}
\begin{split}
\mathrm{Diagram\;}\ref{fig:G^3}\mathrm{(a)}
&= \int dt\,\frac{G^{3}m_1^{2}m_2^{2}}{r^{3}}
\left(
\alpha_1^{3}\alpha_2^{3}
+\alpha_1\alpha_2^{3}\beta_1
+\alpha_1^{3}\alpha_2\beta_2
\right. \\
&\left.\qquad
+\alpha_1\alpha_2\beta_1\beta_2
\right).
\end{split}
\end{flalign}

\begin{flalign}
\begin{split}
	\mathrm{Diagram \; } \ref{fig:G^3} \mathrm{(b)}
	&= \int dt\,\frac{G^{3}m_1^{2}m_2^{2}\alpha_1\alpha_2}{r^{3}},
\end{split}
\end{flalign}

\begin{flalign}
\begin{split}
	\mathrm{Diagram \; } \ref{fig:G^3} \mathrm{(c)}
	&= \int dt\,\frac{G^{3}m_1^{2}m_2^{2}}{r^{3}}\left(\alpha_1^{2}\alpha_2^{2}+\alpha_1^{2}\beta_2\right),
\end{split}
\end{flalign}

\begin{flalign}
\begin{split}
	\mathrm{Diagram \; } \ref{fig:G^3} \mathrm{(d)} &= \int dt\,\frac{G^{3}m_1^{2}m_2^{2}\alpha_1^{2}\alpha_2^{2}}{r^{3}}
\end{split}
\end{flalign}

\begin{flalign}
\begin{split}
	\mathrm{Diagram \; } \ref{fig:G^3} \mathrm{(e)} &= \int dt\,\frac{G^{3}m_1^{2}m_2^{2}\alpha_1\alpha_2}{r^{3}},
\end{split}
\end{flalign}

\begin{flalign}
\begin{split}
	\mathrm{Diagram \; } \ref{fig:G^3} \mathrm{(f)}
	&= \int dt\,\frac{G^3m_1^3m_2\alpha_1^3\alpha_2}{3r^3}.
\end{split}
\end{flalign}

\begin{flalign}
\begin{split}
	\mathrm{Diagram \; } \ref{fig:G^3} \mathrm{(g)}
	&= \int dt\,\frac{G^{3}\,m_1^{3}\,m_2\,\alpha_1\alpha_2}{3\,r^{3}}.
\end{split}
\end{flalign}

\begin{flalign}
\begin{split}
	\mathrm{Diagram \; } \ref{fig:G^3} \mathrm{(h)}
	&=\int dt\,\frac{G^3m_1m_2^3q_2^2\alpha_2^2}{3r^3}
\end{split}
\end{flalign}

\begin{flalign}
\begin{split}
	\mathrm{Diagram \; } \ref{fig:G^3} \mathrm{(i)}
	&= \int dt\,\frac{G^{3}\,m_1m_2^{3}}{2 r^{3}}\alpha_1\alpha_2
\end{split}
\end{flalign}

\begin{flalign}
\begin{split}
	\mathrm{Diagram \; } \ref{fig:G^3} \mathrm{(j)}
	&= \int dt\,\frac{G^3 m_1 m_2^3}{r^3}\left[\frac{1}{6}\beta'_1\alpha_2^3+\frac{1}{6}\alpha_1^3\alpha_2^3+\frac{1}{2}\alpha_1\alpha_2^3\beta_1\right]
\end{split}
\end{flalign}

\begin{flalign}
\begin{split}
	\mathrm{Diagram \; } \ref{fig:G^3} \mathrm{(k)}
	&= \int dt\,\frac{G^{3}\,m_1m_2^{3}}{r^{3}}\left(\frac{1}{2}\alpha_1^{2}\alpha_2^{2}+\frac{1}{2}\alpha_2^{2}\beta_1\right)
\end{split}
\end{flalign}

\begin{flalign}
\begin{split}
	\mathrm{Diagram \; } \ref{fig:G^3} \mathrm{(l)}
	&= \int dt\,\frac{2\,G^{3}\,m_1^{2}m_2^{2}\,\alpha_1^{2}\alpha_2^{2}}{r^{3}}.
\end{split}
\end{flalign}

\begin{flalign}
\begin{split}
	\mathrm{Diagram \; } \ref{fig:G^3} \mathrm{(m)}
	&= \int dt\,\frac{4\,G^{3}\,m_1^{2}\,m_2^{2}\,\alpha_1\alpha_2}{r^{3}}
\end{split}
\end{flalign}

\begin{flalign}
\begin{split}
	\mathrm{Diagram \; } \ref{fig:G^3} \mathrm{(n)}
	&= 0
\end{split}
\end{flalign}

\begin{flalign}
\begin{split}
	\mathrm{Diagram \; } \ref{fig:G^3} \mathrm{(o)}
	&= 0.
\end{split}
\end{flalign}

\subsection{$G^2q^2$ order diagrams}
\label{subsection:G^2q^2}
\begin{figure*}[htbp] 
   \centering
   \includegraphics{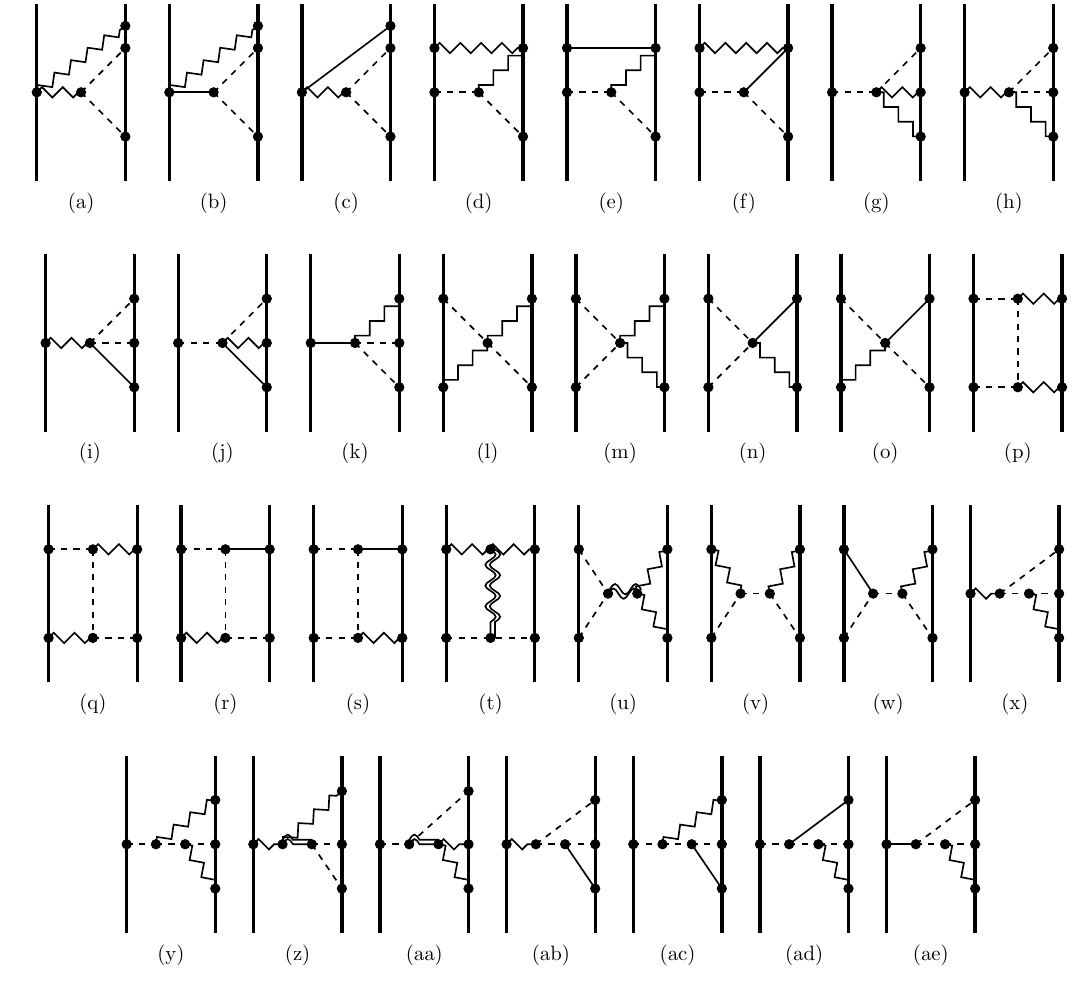}
\caption{The 31 diagrams contributing at $\mathcal{O}(G^2q^2)$. Diagrams (a), (b), (c), (g), (h), (i), (k), (n), (p), (s), (z), and (aa) have a symmetry factor of $1/2$, while diagrams (m) and (u) have a symmetry factor of $1/4$. The remaining diagrams have a symmetry factor of unity.}
\label{fig:G^2q^2}  
\end{figure*}
At $\mathcal{O}(G^2q^2)$, there are 31 two-loop diagrams, as shown in Fig.~\ref{fig:G^2q^2}. These diagrams involve factorizable, nested, and irreducible two-loop topologies. The resulting contributions are

\begin{flalign}
\begin{split}
	\mathrm{Diagram \; } \ref{fig:G^2q^2} \mathrm{(a)}
	&= \int dt\,
\frac{a\,G^{2}m_{1}m_{2}q_{2}^{2}\alpha_{2}}
     {2r^{3}}
\left(\alpha_{1}^{2}+\beta_{1}\right)
	\end{split}
\end{flalign}

\begin{flalign}
\begin{split}
	\mathrm{Diagram \; } \ref{fig:G^2q^2} \mathrm{(b)}
	&= \int dt\,\frac{G^{2}\,m_1\,m_2\,q_2^{2}\,\alpha_1\alpha_2}{2\,r^{3}},
\end{split}
\end{flalign}

\begin{equation}
\begin{split}
	\mathrm{Diagram \; } \ref{fig:G^2q^2} \mathrm{(c)}
	&= \int dt\,\frac{a\,G^{2}\,m_1\,m_2\,q_2^{2}\,\alpha_1}{2\,r^{3}}.
\end{split}
\end{equation}

\begin{flalign}
\begin{split}
	\mathrm{Diagram \; } \ref{fig:G^2q^2} \mathrm{(d)}
	&= -\int dt\,\frac{a\,G^{2}\,m_1m_2\,q_1q_2}{r^{3}}
\left(\alpha_1^{2}\alpha_2+\alpha_2\beta_1\right)
\end{split}
\end{flalign}

\begin{flalign}
\begin{split}
	\mathrm{Diagram \; } \ref{fig:G^2q^2} \mathrm{(e)}
	&= -\int dt\,\frac{a \,G^{2}\,m_1m_2\,q_1q_2\,\alpha_1}{r^{3}},
\end{split}
\end{flalign}

\begin{flalign}
\begin{split}
	\mathrm{Diagram \; } \ref{fig:G^2q^2} \mathrm{(f)}
	&= -\int dt\,\frac{G^{2}\,m_1m_2\,q_1q_2\,\alpha_1\alpha_2}{r^{3}}.
\end{split}
\end{flalign}

\begin{flalign}
\begin{split}
	\mathrm{Diagram \; } \ref{fig:G^2q^2} \mathrm{(g)}
	&= \int dt\,\frac{2\,a^{2}G^{2}\,m_1^{2}\,q_1q_2\,\alpha_1^{2}}{3\,r^{3}},
\end{split}
\end{flalign}

\begin{flalign}
\begin{split}
	\mathrm{Diagram \; } \ref{fig:G^2q^2} \mathrm{(h)}
	&= -\int dt\,\frac{a^{2}\,G^{2}\,m_1\,m_2\,q_1^{2}\,\alpha_1\alpha_2}{3\,r^{3}},
\end{split}
\end{flalign}

\begin{flalign}
\begin{split}
	\mathrm{Diagram \; } \ref{fig:G^2q^2} \mathrm{(i)}
	&= -\int dt\,\frac{a\,G^{2}\,m_1m_2\,q_1^{2}\,\alpha_2}{3\,r^{3}},
\end{split}
\end{flalign}

\begin{flalign}
\begin{split}
	\mathrm{Diagram \; } \ref{fig:G^2q^2} \mathrm{(j)}
	&= \int dt\,\frac{4\,a\,G^{2}\,m_1^{2}\,q_1q_2\,\alpha_1}{3\,r^{3}},
\end{split}
\end{flalign}

\begin{flalign}
\begin{split}
	\mathrm{Diagram \; } \ref{fig:G^2q^2} \mathrm{(k)}
	&= -\int dt\,\frac{a\,G^{2}\,m_1m_2\,q_2^{2}\,\alpha_2}{3\,r^{3}},
\end{split}
\end{flalign}

\begin{flalign}
\begin{split}
	\mathrm{Diagram \; } \ref{fig:G^2q^2} \mathrm{(l)}
	&= 0,
\end{split}
\end{flalign}

\begin{flalign}
\begin{split}
	\mathrm{Diagram \; } \ref{fig:G^2q^2} \mathrm{(m)}
	&= 0,
\end{split}
\end{flalign}

\begin{flalign}
\begin{split}
	\mathrm{Diagram \; } \ref{fig:G^2q^2} \mathrm{(n)}
	&= 0,
\end{split}
\end{flalign}

\begin{flalign}
\begin{split}
	\mathrm{Diagram \; } \ref{fig:G^2q^2} \mathrm{(o)}
	&= 0.
\end{split}
\end{flalign}

\begin{flalign}
\begin{split}
	\mathrm{Diagram \; } \ref{fig:G^2q^2} \mathrm{(p)}
	&= \int dt\,\frac{a^{2}G^{2}\,m_2^{2}\,q_1^{2}\,\alpha_2^{2}}{r^{3}}
\end{split}
\end{flalign}

\begin{flalign}
\begin{split}
	\mathrm{Diagram \; } \ref{fig:G^2q^2} \mathrm{(q)}
	&= \int dt\,\frac{10\,a^{2}G^{2}\,m_1m_2\,q_1q_2\,\alpha_1\alpha_2}{r^{3}}
\end{split}
\end{flalign}

\begin{flalign}
\begin{split}
	\mathrm{Diagram \; } \ref{fig:G^2q^2} \mathrm{(r)}
	&= -\int dt\,\frac{2a\,G^{2}m_1m_2q_1q_2\alpha_2}{r^{3}}
\end{split}
\end{flalign}

\begin{flalign}
\begin{split}
	\mathrm{Diagram \; } \ref{fig:G^2q^2} \mathrm{(s)}
	&= \int dt\,\frac{2\,a\,G^{2}\,m_1^{2}\,q_2^{2}\,\alpha_1}{r^{3}}
\end{split}
\end{flalign}

\begin{flalign}
\begin{split}
	\mathrm{Diagram \; } \ref{fig:G^2q^2} \mathrm{(t)}
	&= -\int dt\,\frac{4G^{2}m_1m_2q_1q_2\alpha_1\alpha_2}{r^{3}}
\end{split}
\end{flalign}

\begin{flalign}
\begin{split}
	\mathrm{Diagram \; } \ref{fig:G^2q^2} \mathrm{(u)}
	&= 0,
\end{split}
\end{flalign}

\begin{flalign}
\begin{split}
	\mathrm{Diagram \; } \ref{fig:G^2q^2} \mathrm{(v)}
	&= 0
\end{split}
\end{flalign}

\begin{flalign}
\begin{split}
	\mathrm{Diagram \; } \ref{fig:G^2q^2} \mathrm{(w)}
	&= 0
\end{split}
\end{flalign}

\begin{flalign}
\begin{split}
	\mathrm{Diagram \; } \ref{fig:G^2q^2} \mathrm{(x)}
	&= \int dt\,\frac{2\,a^{2}\,G^{2}\,m_1\,m_2\,q_1^{2}\,\alpha_1\alpha_2}{3\,r^{3}}
\end{split}
\end{flalign}

\begin{flalign}
\begin{split}
	\mathrm{Diagram \; } \ref{fig:G^2q^2} \mathrm{(y)}
	&= -\int dt\,\frac{4\,a^{2}G^{2}\,m_1^{2}\,q_1q_2\,\alpha_1^{2}}{3\,r^{3}}
\end{split}
\end{flalign}

\begin{flalign}
\begin{split}
	\mathrm{Diagram \; } \ref{fig:G^2q^2} \mathrm{(z)}
	&= -\int dt\,\frac{G^{2}\,m_1\,m_2\,q_1^{2}\,\alpha_1\alpha_2}{3\,r^{3}}.
\end{split}
\end{flalign}

\begin{flalign}
\begin{split}
	\mathrm{Diagram \; } \ref{fig:G^2q^2} \mathrm{(aa)}
	&= -\int dt\,\frac{G^{2}\,m_1^{2}\,q_1q_2\,\alpha_1^{2}}{3\,r^{3}}
\end{split}
\end{flalign}

\begin{flalign}
\begin{split}
	\mathrm{Diagram \; } \ref{fig:G^2q^2} \mathrm{(ab)}
	&= \int dt\,\frac{2\,a\,G^{2}\,m_1\,m_2\,q_1^{2}\,\alpha_2}{3\,r^{3}}
\end{split}
\end{flalign}

\begin{flalign}
\begin{split}
	\mathrm{Diagram \; } \ref{fig:G^2q^2} \mathrm{(ac)}
	&= -\int dt\,\frac{4\,a\,G^{2}\,m_1^{2}\,q_1q_2\,\alpha_1}{3\,r^{3}}
\end{split}
\end{flalign}

\begin{flalign}
\begin{split}
	\mathrm{Diagram \; } \ref{fig:G^2q^2} \mathrm{(ad)}
	&=-\int dt\,\frac{4aG^2m_1^2q_1q_2\alpha_1}{3r^3}
\end{split}
\end{flalign}

\begin{flalign}
\begin{split}
	\mathrm{Diagram \; } \ref{fig:G^2q^2} \mathrm{(ae)}
	&=\int dt\,\frac{2aG^2m_1m_2q_2^2\alpha_2}{3r^3}
\end{split}
\end{flalign}
\\

\subsection{$Gq^4$ order diagrams}
\label{subsection:Gq^4}
\begin{figure}[h]
    \centering
\includegraphics{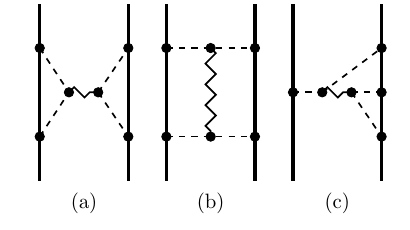} 
\caption{The three diagrams contributing at $\mathcal{O}(Gq^4)$. Diagram (a) has a symmetry factor of $1/4$, while diagrams (b) and (c) have a symmetry factor of $1/2$.}
\label{fig:Gq^4}
\end{figure}
At $\mathcal{O}(Gq^4)$, there are three two-loop diagrams, as shown in Fig.~\ref{fig:Gq^4}. The resulting contributions are

\begin{flalign}
\begin{split}
	\mathrm{Diagram \; } \ref{fig:Gq^4} \mathrm{(a)}
	&= 0.
\end{split}
\end{flalign}

\begin{flalign}
\begin{split}
	\mathrm{Diagram \; } \ref{fig:Gq^4} \mathrm{(b)}
	&= \int dt\,\frac{a^{2} G q_{1}^{2} q_{2}^{2}}{r^{3}}
\end{split}
\end{flalign}

\begin{flalign}
\begin{split}
	\mathrm{Diagram \; } \ref{fig:Gq^4} \mathrm{(c)}
	&= -\int dt\,\frac{a^2Gq_1^3q_2}{3r^3}
\end{split}
\end{flalign}

\section{Results}
\label{sec:results}

Having evaluated all diagrams through 2PN order, we now combine their contributions to obtain the conservative two-body Lagrangian. The velocity-dependent free-particle contribution is obtained by expanding the point-particle action, Eq.~\eqref{eq:sppkk}, through $\mathcal{O}(v^6)$ with the fields set to zero, while the interaction terms are obtained by summing the diagrammatic contributions computed in the preceding sections. In the expression below, $(1\leftrightarrow2)$ denotes the contribution obtained by interchanging the particle labels $1\leftrightarrow2$, together with $\bm n\rightarrow-\bm n$. The complete 2PN Lagrangian is
\begin{widetext}
\begin{align}
\mathcal{L}_{2\rm PN}
&=\tfrac{1}{16}m_1\bm v_1^6
\;+\;\frac{q_1q_2}{16\,r}\bigg\{
   2(\bm v_1\!\cdot\!\bm v_2)^2-\bm v_1^2\bm v_2^2
   +2\bm v_2^2(\bm n\!\cdot\!\bm v_1)^2
   -3(\bm n\!\cdot\!\bm v_1)^2(\bm n\!\cdot\!\bm v_2)^2
\nonumber\\
&\hspace{26mm}
   +r\Big[-4(\bm v_1\!\cdot\!\bm a_2)(\bm n\!\cdot\!\bm v_1)
      -2\bm v_2^2(\bm n\!\cdot\!\bm a_1)
      +2(\bm n\!\cdot\!\bm a_1)(\bm n\!\cdot\!\bm v_2)^2\Big]
   +r^2\Big[-3\,\bm a_1\!\cdot\!\bm a_2
      +(\bm n\!\cdot\!\bm a_1)(\bm n\!\cdot\!\bm a_2)\Big]\bigg\}
\nonumber\\[1.5mm]
&+\frac{G_{12}m_1m_2}{16\,r}\bigg\{
   2\big(7+4\bar\gamma_{12}\big)\bm v_1^4
   -4\big(5+2\bar\gamma_{12}\big)\bm v_1^2(\bm v_1\!\cdot\!\bm v_2)
   +2(\bm v_1\!\cdot\!\bm v_2)^2+3\bm v_1^2\bm v_2^2
   -2\bm v_2^2(\bm n\!\cdot\!\bm v_1)^2
\nonumber\\
&\hspace{28mm}
   -4\big(3+2\bar\gamma_{12}\big)\bm v_1^2
     (\bm n\!\cdot\!\bm v_1)(\bm n\!\cdot\!\bm v_2)
   +4\big(3+2\bar\gamma_{12}\big)(\bm v_1\!\cdot\!\bm v_2)
     (\bm n\!\cdot\!\bm v_1)(\bm n\!\cdot\!\bm v_2)
   +3(\bm n\!\cdot\!\bm v_1)^2(\bm n\!\cdot\!\bm v_2)^2
\nonumber\\
&\hspace{28mm}
   +r\Big[2\big(7+4\bar\gamma_{12}\big)
       \Big((\bm v_1\!\cdot\!\bm a_2)(\bm n\!\cdot\!\bm v_1)
          -(\bm a_1\!\cdot\!\bm v_2)(\bm n\!\cdot\!\bm v_2)\Big)
     +4\big(3+2\bar\gamma_{12}\big)
       \Big((\bm v_1\!\cdot\!\bm a_1)(\bm n\!\cdot\!\bm v_2)
          -(\bm v_2\!\cdot\!\bm a_2)(\bm n\!\cdot\!\bm v_1)\Big)
\nonumber\\
&\hspace{34mm}
     +\bm v_2^2(\bm n\!\cdot\!\bm a_1)-\bm v_1^2(\bm n\!\cdot\!\bm a_2)
     -(\bm n\!\cdot\!\bm a_1)(\bm n\!\cdot\!\bm v_2)^2
     +(\bm n\!\cdot\!\bm v_1)^2(\bm n\!\cdot\!\bm a_2)\Big]
\nonumber\\
&\hspace{28mm}
   +r^2\Big[\big(15+8\bar\gamma_{12}\big)\bm a_1\!\cdot\!\bm a_2
     -(\bm n\!\cdot\!\bm a_1)(\bm n\!\cdot\!\bm a_2)\Big]\bigg\}
\nonumber\\[1.5mm]
&+\frac{Gm_1}{r^2}\bigg\{
   -\tfrac12 q_1q_2\big(1+a\alpha_1\big)\bm v_1^2
   -\tfrac14 q_2^2\big(1-a\alpha_1\big)\bm v_1^2
   +\tfrac12 q_2^2\big(1-a\alpha_1\big)\bm v_1\!\cdot\!\bm v_2
   +\tfrac12 a\,q_2^2\alpha_1\,\bm v_2^2
   +\Big[q_1q_2\big(1+a\alpha_1\big)-\tfrac12 q_2^2\Big]
     (\bm n\!\cdot\!\bm v_1)^2
\nonumber\\
&\hspace{16mm}
   +\Big[q_2^2\big(2+a\alpha_1\big)-2q_1q_2\big(1+a\alpha_1\big)\Big]
     (\bm n\!\cdot\!\bm v_1)(\bm n\!\cdot\!\bm v_2)
   -q_2^2\Big(1+\tfrac12 a\alpha_1\Big)(\bm n\!\cdot\!\bm v_2)^2\bigg\}
\nonumber\\[1.5mm]
&+\frac{G_{12}^2m_1m_2^2}{4r^2}\bigg\{
   \Lambda_1\Big[\bm v_1^2-2\,\bm v_1\!\cdot\!\bm v_2\Big]
   +\Lambda_2\,\bm v_2^2
   +2\Big(1+\bar\gamma_{12}+\tfrac14\bar\gamma_{12}^2+\delta_2\Big)
     (\bm n\!\cdot\!\bm v_1)^2
\nonumber\\
&\hspace{22mm}
   -\Big(4\bar\gamma_{12}+\bar\gamma_{12}^2+4\delta_2-8\bar\beta_1\Big)
     (\bm n\!\cdot\!\bm v_1)(\bm n\!\cdot\!\bm v_2)
   +\Big(2\bar\gamma_{12}+\tfrac12\bar\gamma_{12}^2+2\delta_2
         -4\bar\beta_1\Big)(\bm n\!\cdot\!\bm v_2)^2\bigg\}
\nonumber\\[1.5mm]
&+\frac{G}{r^3}\Big[\tfrac32 q_1^2q_2^2+\tfrac12 a^2q_1^2q_2^2
   -\tfrac13 a^2q_1^3q_2\Big]
\;+\;\frac{G^2m_1^2}{r^3}\bigg\{-q_1q_2+q_2^2
   -\tfrac43 a\,q_1q_2\alpha_1+2a\,q_2^2\alpha_1
   -\tfrac13 q_1q_2\alpha_1^2
   -\tfrac23 a^2q_1q_2\alpha_1^2+a^2q_2^2\alpha_1^2\bigg\}
\nonumber\\[1.5mm]
&+\frac{G^2m_1m_2}{r^3}\bigg\{
   \tfrac12 q_1^2-10\,q_1q_2
   +\tfrac13 a\,q_1^2\alpha_1+\tfrac56 a\,q_2^2\alpha_1
   -3a\,q_1q_2\alpha_1
   +\tfrac16 q_1^2\alpha_1\alpha_2
   +\tfrac13 a^2q_1^2\alpha_1\alpha_2
   -3\,q_1q_2\alpha_1\alpha_2
\nonumber\\
&\hspace{20mm}
   +5\,a^2q_1q_2\alpha_1\alpha_2
   +\tfrac12 a\,q_2^2\alpha_1^2\alpha_2
   -a\,q_1q_2\alpha_1^2\alpha_2
   +\tfrac12 a\,q_2^2\alpha_2\beta_1
   -a\,q_1q_2\alpha_2\beta_1\bigg\}
\nonumber\\[1.5mm]
&+\frac{G_{12}^3m_1m_2^3}{2r^3}
   \Big(1+\tfrac23\bar\gamma_{12}+\tfrac16\bar\gamma_{12}^2
        +2\bar\beta_1+\tfrac23\delta_2+\tfrac13\epsilon_1\Big)
 +\frac{G_{12}^3m_1^2m_2^2}{2r^3}
   \Big(3+\bar\gamma_{12}+4\bar\beta_1+\zeta_{12}\Big)
\;+\;(1\leftrightarrow2),
\label{eq:finalL2PN}
\end{align}
\end{widetext}

where
\begin{align}
\Lambda_1
&\equiv
7+10\bar\gamma_{12}+\frac{7}{2}\bar\gamma_{12}^2
-2\delta_2+2\bar\beta_1,
\\
\Lambda_2
&\equiv
8+10\bar\gamma_{12}+\frac{7}{2}\bar\gamma_{12}^2
-2\delta_2+4\bar\beta_1,
\\
\delta_1
&=\frac{\alpha_1^2}{(1+\alpha_1\alpha_2)^2},
\qquad
\epsilon_1
=\frac{\beta'_1\alpha_2^3}{(1+\alpha_1\alpha_2)^3},
\\
\zeta_{12}
&=\frac{\beta_1\beta_2\alpha_1\alpha_2}
{(1+\alpha_1\alpha_2)^3},
\end{align}
As consistency checks, the 2PN Lagrangian reproduces the known results in the appropriate limits. Setting the scalar couplings to zero, $\alpha_A=\beta_A=\beta'_A=0$, together with $a=0$, reduces the result to the 2PN Einstein-Maxwell Lagrangian of Ref.~\cite{Gupta:2022spq}. In the electrically neutral limit, $q_A=0$, it reduces to the corresponding 2PN scalar-tensor Lagrangian~\cite{Almeida:2024uph}. Setting both the electric and scalar couplings to zero recovers the 2PN conservative dynamics of general relativity~\cite{Gilmore:2008gq}. As an independent check, our result agrees with the corresponding calculation performed using the EFTofPNG package~\cite{Levi:2017kzq}. Finally, in the $G\rightarrow0$ limit, the purely electromagnetic sector agrees with the known 2PN electromagnetic two-body dynamics~\cite{Barker:1980kef}.

\subsection{Static test-body check}
\label{sec:test-body-check}

The test-body limit provides an independent validation of the static
sector of our conservative dynamics. We treat body~1 as an exact EMd
black hole and body~2 as a skeletonized point particle moving in its
background. In harmonic coordinates, the static energy of body~2 is
\begin{equation}
V_{\mathrm{TB}}(r)
=
c^2\mathfrak m_2\!\left[\varphi_1(r)\right]
\sqrt{-g_{00}^{(1)}(r)}
-q_2A_0^{(1)}(r),
\label{eq:V-TB-exact}
\end{equation}
where \(r\) denotes the harmonic radial coordinate. The exact EMd
background, its transformation to harmonic coordinates, and the
intermediate expansion of Eq.~\eqref{eq:V-TB-exact} are presented in
Appendix~\ref{app:test-body-derivation}.

To express the exact result in terms of the scalar-response parameters
used in the two-body calculation, we impose the source black-hole
relation
\begin{equation}
\frac{q_1^2}{Gm_1^2}
=
\frac{2\alpha_1}{a}
-\frac{1-a^2}{a^2}\alpha_1^2.
\label{eq:TB-BH-charge-relation}
\end{equation}
This relation allows us to organize the exact result into the same
$Gq^4$, $G^2m^2q^2$, and $G_{12}^3m^4$ sectors appearing in the
two-body Lagrangian. Expanding through static 2PN order then gives
\begin{widetext}
\begin{align}
V_{\mathrm{TB}}-m_2c^2
={}&
\frac{-G_{12}m_1m_2+q_1q_2}{r}
\nonumber\\[1mm]
&+\frac{1}{c^2r^2}
\bigg[
   -Gm_1q_1q_2\left(1+a\alpha_1\right)
   +\frac{Gm_2q_1^2}{2}\left(1+a\alpha_2\right)
   +\frac{G_{12}^2m_1^2m_2}{2}
      \left(1+2\bar\beta_2\right)
\bigg]
\nonumber\\[1mm]
&+\frac{1}{c^4r^3}
\bigg\{
   \frac{a^2Gq_1^3q_2}{3}
\nonumber\\
&\hspace{13mm}
   +G^2m_1^2q_1q_2
    \left(
       1+\frac{4}{3}a\alpha_1
       +\frac{1}{3}\alpha_1^2
       +\frac{2}{3}a^2\alpha_1^2
    \right)
\nonumber\\
&\hspace{13mm}
   -G^2m_1m_2q_1^2
    \bigg(
       \frac{1}{2}
       +\frac{1}{3}a\alpha_1
       +\frac{5}{6}a\alpha_2
       +\frac{1}{6}\alpha_1\alpha_2
       +\frac{1}{3}a^2\alpha_1\alpha_2
       +\frac{1}{2}a\alpha_1\alpha_2^2
       +\frac{1}{2}a\alpha_1\beta_2
    \bigg)
\nonumber\\
&\hspace{13mm}
   -\frac{G_{12}^3m_1^3m_2}{2}
    \left(
       1+\frac{2}{3}\bar\gamma_{12}
       +\frac{1}{6}\bar\gamma_{12}^{\,2}
       +2\bar\beta_2
       +\frac{2}{3}\delta_1
       +\frac{1}{3}\epsilon_2
    \right)
\bigg\}
+O(c^{-6}).
\label{eq:V-TB-2PN}
\end{align}
\end{widetext}
The static interaction Lagrangian is related to the test-body energy by
\begin{equation}
\mathcal L_{\mathrm{TB,int}}^{\mathrm{stat}}
=
-\left(V_{\mathrm{TB}}-m_2c^2\right).
\label{eq:LminusV-TB}
\end{equation}

We take the test-body limit of the conservative two-body Lagrangian by
rescaling
\[
m_2\rightarrow\varepsilon m_2,
\qquad
q_2\rightarrow\varepsilon q_2,
\qquad
\varepsilon\rightarrow0,
\]
while keeping $q_2/m_2$, $\alpha_2$, $\beta_2$, and $\beta'_2$ fixed.
The test-body contribution is obtained by retaining terms linear in
$\varepsilon$. We then restrict to the static sector by setting
$\bm v_A=\bm a_A=0$ and include the contribution obtained under
$(1\leftrightarrow2)$. We find
\begin{equation}
\left.
\left(
\mathcal L_{\mathrm{0PN}}
+\mathcal L_{\mathrm{1PN}}
+\mathcal L_{\mathrm{2PN}}
\right)
\right|_{\mathrm{stat,TB}}
=
\mathcal L_{\mathrm{TB,int}}^{\mathrm{stat}}
+O(c^{-6}).
\label{eq:TB-static-match}
\end{equation}
Thus, the static test-body limit of the conservative two-body
Lagrangian exactly reproduces the interaction obtained independently
from the exact EMd black-hole background through 2PN order. This
provides a nontrivial independent check of the static sector of our
2PN result Eq.~\eqref{eq:finalL2PN}.

\section{Conclusions}
\label{sec:conclusions}
In this work, we have derived the conservative two-body dynamics of charged black-hole binaries in EMd theory through 2PN order using the EFT approach. We derived the worldline and bulk Feynman rules required at this order and evaluated the diagrams contributing to the conservative effective Lagrangian. The 2PN calculation involves the $\mathcal{O}(Gv^4)$, $\mathcal{O}(G^2v^2)$, $\mathcal{O}(Gq^2v^2)$, $\mathcal{O}(G^3)$, $\mathcal{O}(G^2q^2)$, and $\mathcal{O}(Gq^4)$ sectors, including one- and two-loop diagrams. The two-loop contributions can be organized into factorizable, nested, and irreducible topologies and evaluated systematically within the EFT framework.

Our result extends the conservative dynamics of EMd binaries beyond the previously known 1PN order and provides the conservative Lagrangian through 2PN order, including the effects of the electric charges, scalar responses, and the dilaton coupling $a$. The result reproduces the known Einstein-Maxwell, scalar-tensor, and general-relativistic dynamics in
the appropriate limits. As an independent check, we considered the test-body limit, treating one body as a point particle moving in the exact EMd black-hole background. The weak-field expansion of the exact solution reproduces the static test-body limit of the two-body Lagrangian through 2PN order, providing a nontrivial validation of the static sector of our result.

The conservative dynamics obtained here provide a basis for extending the description of EMd binaries toward gravitational-wave observables. A complete waveform model also requires the corresponding radiative sector, including scalar, electromagnetic, and gravitational radiation. In particular, the additional scalar and electromagnetic degrees of freedom can introduce dissipative effects that enter earlier in the PN expansion than the leading gravitational-wave radiation reaction in general relativity. Such effects lie beyond the conservative near-zone dynamics considered in the present work.

Several extensions of the present calculation are possible. The 2PN dynamics can be used to derive the equations of motion and gauge-invariant quantities for bound binaries and to develop an effective-one-body description of the conservative dynamics. Combined with a PN-consistent treatment of the radiative sector, these results can be used to construct inspiral waveforms and investigate the observational signatures of charged black-hole binaries in EMd theory. Extensions to higher PN orders and the inclusion of finite-size effects provide further directions for future work.



\acknowledgements
The author thanks Micha{\l} Bejger for valuable comments that helped improve the manuscript, and is grateful to Jan Steinhoff and Tanja Hinderer for valuable discussions. This work was partially supported by the Polish National Science Centre (NCN) through Grant No.~2021/43/B/ST9/01714.

\appendix

\section{Derivation of the static test-body limit}
\label{app:test-body-derivation}

In this appendix we derive the static test-body potential used in Sec.~\ref{sec:test-body-check}.  We set the asymptotic scalar field to zero.

\subsection{Exact background in harmonic coordinates}

We denote the radial coordinate of the exact electrically charged EMd
black-hole solution by $\rho$.  In the Einstein frame, the metric is
\begin{equation}
\mathrm{d}s^{2}
=-A(\rho)\,\mathrm{d}t^{2}
+A^{-1}(\rho)\,\mathrm{d}\rho^{2}
+\rho^{2}C(\rho)\,\mathrm{d}\Omega^{2},
\label{eq:ghs-metric-app}
\end{equation}
where
\begin{align}
A(\rho)
&=\left(1-\frac{r_{+}}{\rho}\right)
  \left(1-\frac{r_{-}}{\rho}\right)^{\gamma},
&
C(\rho)
&=\left(1-\frac{r_{-}}{\rho}\right)^{1-\gamma},
\label{eq:ghs-functions-app}
\\
\gamma&\equiv\frac{1-a^{2}}{1+a^{2}}.
\end{align}
The mass, electric charge, and dilaton charge satisfy
\begin{align}
2GM&=r_{+}+\gamma r_{-},
&
GQ^{2}&=\frac{r_{+}r_{-}}{1+a^{2}},
&
GD&=\frac{a r_{-}}{1+a^{2}}.
\label{eq:background-relations-app}
\end{align}
Consequently, for $a\neq0$,
\begin{align}
r_{-}&=G\frac{1+a^{2}}{a}D,
&
r_{+}&=2GM-G\frac{1-a^{2}}{a}D,
\label{eq:rpm-app}
\\
Q^{2}
&=G\left[
\frac{2MD}{a}-\frac{1-a^{2}}{a^{2}}D^{2}
\right].
\label{eq:QMD-app}
\end{align}
The scalar and electrostatic fields are
\begin{equation}
\varphi(\rho)
=\frac{a}{1+a^{2}}
 \ln\left(1-\frac{r_{-}}{\rho}\right),
\qquad
A_{0}(\rho)=-\frac{Q}{\rho}.
\label{eq:fields-rho-app}
\end{equation}

Let $x^{i}=r(\rho)n^{i}$ be harmonic Cartesian coordinates, with
$n^{i}n^{i}=1$.  The condition $\Box_{g}x^{i}=0$ reduces to
\begin{equation}
\frac{\mathrm{d}}{\mathrm{d}\rho}
\left[
\rho^{2}A(\rho)C(\rho)\frac{\mathrm{d}r}{\mathrm{d}\rho}
\right]-2r=0.
\label{eq:harmonic-condition-app}
\end{equation}
Since
\begin{equation}
\rho^{2}A(\rho)C(\rho)
=(\rho-r_{+})(\rho-r_{-}),
\end{equation}
the asymptotically Cartesian branch is simply
\begin{equation}
r=\rho-\frac{r_{+}+r_{-}}{2}.
\label{eq:harmonic-radius-app}
\end{equation}
It is useful to introduce
\begin{equation}
\Sigma\equiv\frac{r_{+}+r_{-}}{2}=G(M+aD),
\qquad
\Delta\equiv\frac{r_{+}-r_{-}}{2}
=G\left(M-\frac{D}{a}\right),
\label{eq:SigmaDelta-app}
\end{equation}
so that $\rho=r+\Sigma$.  The temporal component of the exact metric then
becomes
\begin{equation}
-g_{00}^{\mathrm H}(r)
=\frac{
\left(1-\dfrac{\Delta}{r}\right)
\left(1+\dfrac{\Delta}{r}\right)^{\gamma}
}{
\left(1+\dfrac{\Sigma}{r}\right)^{1+\gamma}
}.
\label{eq:g00-harmonic-app}
\end{equation}
The exact metric consequently reduces smoothly to the harmonic-coordinate
Reissner-Nordstr\"om form and harmonic-coordinate Schwarzschild
\begin{equation}
-g_{00}^{\mathrm H}
\xrightarrow[a\to0]{}
\frac{1-\dfrac{G^{2}M^{2}-GQ^{2}}{r^{2}}}
{\left(1+\dfrac{GM}{r}\right)^{2}},
\label{eq:RN-harmonic-app}
\end{equation}

\begin{equation}
-g_{00}^{\mathrm H}
\xrightarrow[a\to0, q\to 0]{}
\frac{1-GM/r}{1+GM/r}
\label{eq:RN-harmonic-app}
\end{equation}

For completeness, the full line element is
\begin{align}
\mathrm{d}s^{2}
={}&-A_{\mathrm H}(r)\,\mathrm{d}t^{2}
+A_{\mathrm H}^{-1}(r)\,\mathrm{d}r^{2}
\nonumber\\
&+(r+\Sigma)^{2}
\left(\frac{r+\Delta}{r+\Sigma}\right)^{\frac{2a^{2}}{1+a^{2}}}
\mathrm{d}\Omega^{2},
\label{eq:full-harmonic-metric-app}
\end{align}
where $A_{\mathrm H}=-g_{00}^{\mathrm H}$.  The background fields in the
same coordinates are
\begin{align}
\delta\varphi_{1}(r)
&\equiv\varphi_{1}(r)-\varphi_{0}
=\frac{a}{1+a^{2}}
\ln\left(\frac{1+\Delta/r}{1+\Sigma/r}\right),
\label{eq:scalar-harmonic-app}
\\
A_{0,1}(r)&=-\frac{Q}{r+\Sigma}.
\label{eq:A0-harmonic-app}
\end{align}
Thus the radial coordinate entering the PN comparison is the harmonic
separation $r$, rather than the original coordinate $\rho$.

\subsection{Weak-field expansion}

For a static test body with scalar-dependent mass
$\mathfrak m_{2}(\varphi)$ and charge $q_{2}$, the exact energy is
\begin{equation}
V_{\mathrm{TB}}
=c^2\mathfrak m_{2}\!\left[\varphi_{1}(r)\right]
 \sqrt{-g_{00}^{\mathrm H}(r)}
-q_{2}A_{0,1}(r).
\label{eq:Vexact-app}
\end{equation}
The required mass expansion is most conveniently written as
\begin{equation}
\ln\frac{\mathfrak m_{2}(\varphi_{1})}{m_{2}}
=\alpha_{2}\delta\varphi_{1}
+\frac{1}{2}\beta_{2}\delta\varphi_{1}^{2}
+\frac{1}{6}\beta'_{2}\delta\varphi_{1}^{3}
+\mathcal O(\delta\varphi_{1}^{4}).
\label{eq:mass-log-app}
\end{equation}

We define
\begin{equation}
\mathcal S\equiv\Delta^{2}+\Delta\Sigma+\Sigma^{2}.
\label{eq:S-app}
\end{equation}
The following identities are useful:
\begin{align}
a^{2}\Delta+\Sigma&=GM(1+a^{2}),
\nonumber\\
\Sigma^{2}-\Delta^{2}&=GQ^{2}(1+a^{2}),
\nonumber\\
\Delta-\Sigma&=-\frac{GD}{a}(1+a^{2}),
\label{eq:identities-app}
\\
\frac{a^{2}\Delta^{3}+\Sigma^{3}}{1+a^{2}}
&=G^{3}(M^{3}+MD^{2})+aG^{2}DQ^{2}.
\label{eq:cubic-identity-app}
\end{align}
Expanding the scalar field Eq.~\eqref{eq:scalar-harmonic-app}, metric factor Eq.~\eqref{eq:g00-harmonic-app}, and electrostatic potential Eq.~\eqref{eq:A0-harmonic-app}, at large $r$ gives
\begin{align}
\delta\varphi_{1}
={}&-\frac{GD}{r}
+\frac{aGQ^{2}}{2r^{2}}
-\frac{GD\mathcal S}{3r^{3}}
+\mathcal O(r^{-4}),
\label{eq:scalar-series-app}
\\
\ln\sqrt{-g_{00}^{\mathrm H}}
={}&-\frac{GM}{r}
+\frac{GQ^{2}}{2r^{2}}
\nonumber\\
&-\frac{G^{3}(M^{3}+MD^{2})+aG^{2}DQ^{2}}
{3r^{3}}
+\mathcal O(r^{-4}),
\label{eq:metric-series-app}
\\
-q_{2}A_{0,1}
={}&\frac{q_{2}Q}{r}
-\frac{q_{2}Q\Sigma}{r^{2}}
+\frac{q_{2}Q\Sigma^{2}}{r^{3}}
+\mathcal O(r^{-4}).
\label{eq:gauge-series-app}
\end{align}

Combining the logarithmic mass and metric expansions, we write
\begin{equation}
\ln\left[
\frac{\mathfrak m_2(\varphi_1)}{m_2}
\sqrt{-g_{00}^{\rm H}}
\right]
=
\frac{E_1}{r}+\frac{E_2}{r^2}+\frac{E_3}{r^3}
+\mathcal O(r^{-4}),
\end{equation}
which gives
\begin{align}
\frac{\mathfrak m_{2}(\varphi_{1})}{m_{2}}
\sqrt{-g_{00}^{\mathrm H}}
={}&1+\frac{E_{1}}{r}
+\frac{E_{2}+\tfrac12E_{1}^{2}}{r^{2}}
\nonumber\\
&+\frac{E_{3}+E_{1}E_{2}+\tfrac16E_{1}^{3}}{r^{3}}
+\mathcal O(r^{-4}).
\label{eq:exponential-app}
\end{align}
where the coefficients are
\begin{align}
E_{1}
={}&-G(M+\alpha_{2}D),
\label{eq:E1-app}
\\
E_{2}
={}&\frac{1}{2}
\left[G(1+a\alpha_{2})Q^{2}+G^{2}\beta_{2}D^{2}\right],
\label{eq:E2-app}
\\
E_{3}
={}&-\frac{a^{2}\Delta^{3}+\Sigma^{3}}{3(1+a^{2})}
-\frac{\alpha_{2}GD}{3}\mathcal S
\nonumber\\
&-\frac{a\beta_{2}G^{2}DQ^{2}}{2}
-\frac{\beta'_{2}G^{3}D^{3}}{6}.
\label{eq:E3-app}
\end{align}
Therefore,
\begin{align}
V_{\mathrm{TB}}-m_{2}c^2
={}&\frac{m_{2}E_{1}+q_{2}Q}{r}
\nonumber\\
&+\frac{m_{2}\left(E_{2}+\tfrac12E_{1}^{2}\right)
-q_{2}Q\Sigma}{r^{2}}
\nonumber\\
&+\frac{m_{2}\left(E_{3}+E_{1}E_{2}
+\tfrac16E_{1}^{3}\right)+q_{2}Q\Sigma^{2}}{r^{3}}
+\mathcal O(r^{-4}).
\label{eq:Vgeneral-app}
\end{align}

For the source black hole, we identify
\begin{equation}
M=m_1,\qquad
D=m_1\alpha_1,\qquad
Q=q_1.
\label{eq:source-identification-app}
\end{equation}
Using these identifications together with the black-hole charge relation, Eq.~\eqref{eq:TB-BH-charge-relation}, Eq.~\eqref{eq:Vgeneral-app} reduces to the test-body expansion given in Eq.~\eqref{eq:V-TB-2PN}.

\section{Useful formulas}
\label{app:useful-formulas}
We flip the time derivative between the two particles  by using the identity
\begin{widetext}
\begin{align}
    \label{eq:timeflip}
    \int dt_1 dt_2  \partial_{t_1}\delta(t_1-t_2) f(t_1) g(t_2) 
    &= -\int dt_1  dt_2  \partial_{t_2}\delta(t_1-t_2) f(t_1) g(t_2).
\end{align}
\end{widetext}
We evaluate the Fourier integrals that we often encounter from propagators and loop integrals using the d-dimensional master formula given by 
\begin{equation}
\label{eq:ft}
\int \frac{d^d \mathbf{k}}{(2\pi)^d} \frac{e^{i \mathbf{k} \cdot \mathbf{r}}}{(\mathbf{k}^2)^\alpha}
= \frac{1}{(4\pi)^{d/2}} \frac{\Gamma(d/2 - \alpha)}{\Gamma(\alpha)} 
\left(\frac{\mathbf{r}^2}{4}\right)^{\alpha - d/2}.
\end{equation}
From this master formula, we obtain the following required Fourier integrals \cite{Levi:2011eq}:
\begin{widetext}
\begin{eqnarray}
\label{tensorfourierindentity}
I^i & \equiv & \int \frac{d^d \mathbf{k}}{(2\pi)^d} \frac{k^i e^{i \mathbf{k} \cdot \mathbf{r}}}{(\mathbf{k}^2)^\alpha}  =  \frac{i}{(4\pi)^{d/2}} \frac{\Gamma(d/2 - \alpha + 1)}{\Gamma(\alpha)}
\left(\frac{\mathbf{r}^2}{4}\right)^{\alpha - d/2 - 1/2} n^i, \\
I^{ij} & \equiv & \int \frac{d^d \mathbf{k}}{(2\pi)^d} \frac{k^i k^j e^{i \mathbf{k} \cdot \mathbf{r}}}{(\mathbf{k}^2)^\alpha} = \frac{1}{(4\pi)^{d/2}} \frac{\Gamma(d/2 - \alpha + 1)}{\Gamma(\alpha)}
\left(\frac{\mathbf{r}^2}{4}\right)^{\alpha - d/2 - 1}
\left(\frac{1}{2} \delta^{ij} + (\alpha - 1 - d/2) n^i n^j \right), \\
I^{ijl} & \equiv & \int \frac{d^d \mathbf{k}}{(2\pi)^d} \frac{k^i k^j k^l e^{i \mathbf{k} \cdot \mathbf{r}}}{(\mathbf{k}^2)^\alpha} = \frac{i}{(4\pi)^{d/2}} \frac{\Gamma(d/2 - \alpha + 2)}{\Gamma(\alpha)}
\left(\frac{\mathbf{r}^2}{4}\right)^{\alpha - d/2 - 3/2} \nonumber \\
& & \times \left( \frac{1}{2} \left(\delta^{ij} n^l + \delta^{il} n^j + \delta^{jl} n^i \right)
+ (\alpha - d/2 - 2) n^i n^j n^l \right), \\
I^{ijlm} & \equiv & \int \frac{d^d \mathbf{k}}{(2\pi)^d} \frac{k^i k^j k^l k^m e^{i \mathbf{k} \cdot \mathbf{r}}}{(\mathbf{k}^2)^\alpha} = \frac{1}{(4\pi)^{d/2}} \frac{\Gamma(d/2 - \alpha + 2)}{\Gamma(\alpha)}
\left(\frac{\mathbf{r}^2}{4}\right)^{\alpha - d/2 - 2}  \times \left( \frac{1}{4} \left( \delta^{ij} \delta^{lm} + \delta^{il} \delta^{jm} + \delta^{im} \delta^{jl} \right) \right. \nonumber \\
& & + \frac{\alpha - d/2 - 2}{2} \left( \delta^{ij} n^l n^m + \delta^{il} n^j n^m 
+ \delta^{im} n^j n^l + \delta^{jl} n^i n^m + \delta^{jm} n^i n^l + \delta^{lm} n^i n^j \right) \nonumber \\
& & \left. + (\alpha - d/2 - 2)(\alpha - d/2 - 3) n^i n^j n^l n^m \right).
\end{eqnarray}
\end{widetext}

We use the d-dimensional master formula for one-loop scalar integrals given by 
\begin{widetext}
\begin{equation}
    J \equiv \int \frac{d^d \mathbf{k}}{(2\pi)^d} \frac{1}{\left[ \mathbf{k}^2 \right]^\alpha \left[ (\mathbf{k} - \mathbf{q})^2 \right]^\beta} 
    = \frac{1}{(4\pi)^{d/2}} \frac{\Gamma(\alpha + \beta - d/2)}{\Gamma(\alpha) \Gamma(\beta)} 
    \frac{\Gamma(d/2 - \alpha) \Gamma(d/2 - \beta)}{\Gamma(d - \alpha - \beta)} 
    \left( q^2 \right)^{d/2 - \alpha - \beta}.
    \label{eq:1loop}
\end{equation}
\end{widetext}
The d-dimensional master formula for one-loop tensor integrals is also taken from \cite{Levi:2011eq}. 
Similarly, one can also derive the following d-dimensional master formulae for the one-loop tensor integrals: 
\begin{widetext}
\begin{align}
    \label{eq:1loopvec}
    J^i & \equiv \int \frac{d^d \mathbf{k}}{(2\pi)^d} \frac{k^i}{\left[\mathbf{k}^2\right]^\alpha \left[(\mathbf{k} - \mathbf{q})^2\right]^\beta} = \frac{1}{(4\pi)^{d/2}} \frac{\Gamma(\alpha + \beta - d/2)}{\Gamma(\alpha) \Gamma(\beta)} 
    \frac{\Gamma(d/2 - \alpha + 1) \Gamma(d/2 - \beta)}{\Gamma(d - \alpha - \beta + 1)} 
    \left(q^2\right)^{d/2 - \alpha - \beta} q^i, \\
    J^{ij} & \equiv \int \frac{d^d \mathbf{k}}{(2\pi)^d} \frac{k^i k^j}{\left[\mathbf{k}^2\right]^\alpha \left[(\mathbf{k} - \mathbf{q})^2\right]^\beta} = \frac{1}{(4\pi)^{d/2}} \frac{\Gamma(\alpha + \beta - d/2 - 1)}{\Gamma(\alpha) \Gamma(\beta)} 
    \frac{\Gamma(d/2 - \alpha + 1) \Gamma(d/2 - \beta)}{\Gamma(d - \alpha - \beta + 2)} 
    \left(q^2\right)^{d/2 - \alpha - \beta} \nonumber \\
    & \quad \times \left[ \frac{d/2 - \beta}{2} q^2 \delta^{ij} + (\alpha + \beta - d/2 - 1) (d/2 - \alpha + 1) q^i q^j \right], \\
    J^{ijl} & \equiv \int \frac{d^d \mathbf{k}}{(2\pi)^d} \frac{k^i k^j k^l}{\left[\mathbf{k}^2\right]^\alpha \left[(\mathbf{k} - \mathbf{q})^2\right]^\beta} = \frac{1}{(4\pi)^{d/2}} \frac{\Gamma(\alpha + \beta - d/2 - 1)}{\Gamma(\alpha) \Gamma(\beta)} 
    \frac{\Gamma(d/2 - \alpha + 2) \Gamma(d/2 - \beta)}{\Gamma(d - \alpha - \beta + 3)} 
    \left(q^2\right)^{d/2 - \alpha - \beta} \nonumber \\
    & \quad \times \left[ \frac{d/2 - \beta}{2} q^2 \left(\delta^{ij} q^l + \delta^{il} q^j + \delta^{jl} q^i \right) + (\alpha + \beta - d/2 - 1) (d/2 - \alpha + 2) \frac{d/2 - \beta}{2} q^2 \right. \nonumber \\
    & \quad \times \left(\delta^{ij} q^l q^m + \delta^{il} q^j q^m + \delta^{im} q^j q^l + \delta^{jl} q^i q^m + \delta^{jm} q^i q^l + \delta^{lm} q^i q^j \right) \nonumber \\
    & \quad \left. + (\alpha + \beta - d/2 - 2) (\alpha + \beta - d/2 - 1) (d/2 - \alpha + 2) (d/2 - \alpha + 3) q^i q^j q^l q^m \right].
\end{align}
\end{widetext}

In addition, we encounter irreducible two-loop tensor integrals up to order 4. Using an integration by parts method as in \cite{Smirnov:2004ym}, these can be written as a sum of factorizable and nested two-loops, as explained in Sec.~\ref{subsection:Gq^4},\ref{subsection:G^2q^2}. The required irreducible two-loop tensor integral reductions are given by
\begin{widetext}
\begin{align}
    \label{eq:2tensor2loop}
    \int_{\mathbf{k}_1 \mathbf{k}_2} \frac{k_1^i k_2^j}{\bm{k}_1^2 \left(p - k_1\right)^2 \bm{k}_2^2 \left(p - k_2\right)^2 \left(k_1 - k_2\right)^2} 
    &= \frac{1}{d - 3} \int_{\mathbf{k}_1 \mathbf{k}_2} \Bigg[ 
        \frac{p^i k_2^j}{k_1^4 \left(p - k_1\right)^2 \bm{k}_2^2 \left(p - k_2\right)^2} 
        - \frac{k_1^i k_2^j}{k_1^4 \left(p - k_1\right)^2 \left(p - k_2\right)^2 \left(k_1 - k_2\right)^2} \nonumber \\
    & \quad - \frac{k_1^i k_2^j}{\bm{k}_1^2 \left(p - k_1\right)^4 \bm{k}_2^2 \left(k_1 - k_2\right)^2} 
        + \frac{1}{d - 4} \Bigg( 
            2 \frac{k_2^i k_2^j}{k_1^4 \left(p - k_1\right)^2 \bm{k}_2^2 \left(p - k_2\right)^2} 
            \nonumber \\
    & \quad - \frac{k_2^i k_2^j}{k_1^4 \left(p - k_1\right)^2 \left(p - k_2\right)^2 \left(k_1 - k_2\right)^2}  - \frac{k_2^i k_2^j}{\bm{k}_1^2 \left(p - k_1\right)^4 \bm{k}_2^2 \left(k_1 - k_2\right)^2} 
        \Bigg) \Bigg], \\
    \label{eq:3tensor2loop}
    \int_{\mathbf{k}_1 \mathbf{k}_2} \frac{k_1^i k_1^j k_2^l}{\bm{k}_1^2 \left(p - k_1\right)^2 \bm{k}_2^2 \left(p - k_2\right)^2 \left(k_1 - k_2\right)^2} 
    &= \frac{1}{d - 3} \int_{\mathbf{k}_1 \mathbf{k}_2} \Bigg[ 
        \frac{p^l k_1^i k_1^j}{\bm{k}_1^2 \left(p - k_1\right)^2 k_2^4 \left(p - k_2\right)^2} 
        - \frac{k_1^i k_1^j k_2^l}{\left(p - k_1\right)^2 k_2^4 \left(p - k_2\right)^2 \left(k_1 - k_2\right)^2} \nonumber \\
    & \quad - \frac{k_1^i k_1^j k_2^l}{\bm{k}_1^2 \bm{k}_2^2 \left(p - k_2\right)^4 \left(k_1 - k_2\right)^2} 
        + \frac{1}{d - 4} \Bigg( 
            2 \frac{k_1^i k_1^j k_1^l}{\bm{k}_1^2 \left(p - k_1\right)^2 k_2^4 \left(p - k_2\right)^2} \nonumber \\
    & \quad
            - \frac{k_1^i k_1^j k_1^l}{\left(p - k_1\right)^2 k_2^4 \left(p - k_2\right)^2 \left(k_1 - k_2\right)^2}  - \frac{k_1^i k_1^j k_1^l}{\bm{k}_1^2 \bm{k}_2^2 \left(p - k_2\right)^4 \left(k_1 - k_2\right)^2} 
        \Bigg) \Bigg], \\
    \label{eq:4tensor2loop}
    \int_{\mathbf{k}_1 \mathbf{k}_2} \frac{k_1^i k_1^j k_2^l k_2^m}{\bm{k}_1^2 \left(p - k_1\right)^2 \bm{k}_2^2 \left(p - k_2\right)^2 \left(k_1 - k_2\right)^2} 
    &= \frac{1}{d - 2} \int_{\mathbf{k}_1 \mathbf{k}_2} \Bigg[ 
        \frac{k_1^i k_1^j k_2^l k_2^m}{k_1^4 \left(p - k_1\right)^2 \bm{k}_2^2 \left(p - k_2\right)^2} 
        + \frac{k_1^i k_1^j k_2^l k_2^m}{\bm{k}_1^2 \left(p - k_1\right)^4 \bm{k}_2^2 \left(p - k_2\right)^2} \nonumber \\
    & \quad - \frac{k_1^i k_1^j k_2^l k_2^m}{k_1^4 \left(p - k_1\right)^2 \left(p - k_2\right)^2 \left(k_1 - k_2\right)^2} 
        - \frac{k_1^i k_1^j k_2^l k_2^m}{\bm{k}_1^2 \left(p - k_1\right)^4 \bm{k}_2^2 \left(k_1 - k_2\right)^2} \nonumber \\
    & \quad + \frac{1}{d - 3} \Bigg( 
            \frac{p^i k_2^j k_2^l k_2^m}{k_1^4 \left(p - k_1\right)^2 \bm{k}_2^2 \left(p - k_2\right)^2} 
            + \frac{p^j k_2^i k_2^l k_2^m}{k_1^4 \left(p - k_1\right)^2 \bm{k}_2^2 \left(p - k_2\right)^2} \nonumber \\
    & \quad - \frac{k_1^i k_2^j k_2^l k_2^m}{k_1^4 \left(p - k_1\right)^2 \left(p - k_2\right)^2 \left(k_1 - k_2\right)^2} 
            - \frac{k_1^j k_2^i k_2^l k_2^m}{k_1^4 \left(p - k_1\right)^2 \left(p - k_2\right)^2 \left(k_1 - k_2\right)^2} \nonumber \\
    & \quad - \frac{k_1^i k_2^j k_2^l k_2^m}{\bm{k}_1^2 \left(p - k_1\right)^4 \bm{k}_2^2 \left(k_1 - k_2\right)^2} 
            - \frac{k_1^j k_2^i k_2^l k_2^m}{\bm{k}_1^2 \left(p - k_1\right)^4 \bm{k}_2^2 \left(k_1 - k_2\right)^2} \nonumber \\
    & \quad + \frac{2}{d - 4} \Bigg( 
            2 \frac{k_2^i k_2^j k_2^l k_2^m}{k_1^4 \left(p - k_1\right)^2 \bm{k}_2^2 \left(p - k_2\right)^2} 
            - \frac{k_2^i k_2^j k_2^l k_2^m}{k_1^4 \left(p - k_1\right)^2 \left(p - k_2\right)^2 \left(k_1 - k_2\right)^2} \nonumber \\
& \quad - \frac{k_2^i k_2^j k_2^l k_2^m}{\bm{k}_1^2 \left(p - k_1\right)^4 \bm{k}_2^2 \left(k_1 - k_2\right)^2}
\Bigg) \Bigg].
\end{align}
\end{widetext}
It should be noted that these expressions contain explicit poles in $d=3$, but these cancel out in the dimensional regularization. 

\bibliography{main_auto}

\end{document}